\documentclass[11pt,a4paper]{article} 
\pdfoutput=1
\usepackage{jheppub}

\usepackage{epsfig,multicol,bbm}
\usepackage{graphicx}
\usepackage{amsmath}
\usepackage{amssymb}
\usepackage{bm}
\usepackage{multirow}
\usepackage{slashed}
\allowdisplaybreaks[1]

\newcommand{\ET}{\mbox{$\not \hspace{-0.10cm} E_T$ }}
\newcommand\fverb{\setbox\fverbbox=\hbox\bgroup\verb}
\newcommand\fverbdo{\egroup\medskip\noindent%
			\fbox{\unhbox\fverbbox}\ }
\newcommand\fverbit{\egroup\item[\fbox{\unhbox\fverbbox}]}
\newbox\fverbbox

\title{Search for Higgs portal DM at the ILC}

\author[a,b]{P. Ko,}
\author[b]{Hiroshi Yokoya}

\affiliation[a]{School of Physics, KIAS, Seoul 02455, Korea}
\affiliation[b]{Quantum Universe Center, KIAS, Seoul 02455, Korea}

\emailAdd{pko@kias.re.kr}
\emailAdd{hyokoya@kias.re.kr}

\arxivnumber{}	
\abstract{ 
Higgs portal dark matter (DM) models are simple interesting and viable
DM models. 
There are three types of the models depending on the DM spin: scalar,
fermion and vector DM models.
In this paper, we consider renormalizable, unitary and gauge invariant
Higgs portal DM models, and study how large parameter regions can be
surveyed at the International Linear Collider (ILC) experiment at
$\sqrt{s}=500$~GeV.
For the Higgs portal singlet fermion and vector DM cases, the force
mediator involves two scalar propagators, the SM-like Higgs boson and
the dark Higgs boson.
We show that their interference generates interesting and important
patterns in the mono-$Z$ plus missing $E_T$ signatures at the ILC,
and the results are completely different from those obtained from the
Higgs portal DM models within the effective field theories.
In addition, we show that it would be possible to distinguish the spin
of DM in the Higgs portal scenarios, if the shape of the recoil-mass
distribution is observed.
We emphasize that the interplay between these collider observations and
those in the direct detection experiments has to be performed in the
model with renomalizability and unitarity to combine the model analyses
in different scales.
}


\begin{document}
\maketitle
\section{Introduction\label{sec:intro}}

One of the most pressing question after the Higgs boson discovery is to
understand particle physics nature of nonbaryonic dark matter (DM) of
the universe.
Existence of DM has been confirmed only through gravitation probes, and
it is important to find their properties (mass, spin, quantum numbers,
etc.) using terrestrial experimental apparatus.
Among many DM models, Higgs portal scalar, fermion and vector DM
models are simple and interesting~\cite{Kanemura:2010sh,%
Lebedev:2011iq,Djouadi:2011aa,LopezHonorez:2012kv,Endo:2014cca,%
Craig:2014lda,Beniwal:2015sdl,Han:2016gyy}, which are also
phenomenologically viable and have intimate connections to the observed
125~GeV Higgs
boson~\cite{Aad:2012tfa,Chatrchyan:2012xdj,Chatrchyan:2013lba}. 
Study on the characteristic signatures at collider experiments has to be
performed to verify these models.
In particular it would be important to know if one can distinguish the
DM spin at current or future colliders. 

In this work, we present a detailed study on this issue at the
International Linear Collider (ILC)~\cite{Baer:2013cma,Fujii:2015jha,%
Barklow:2015tja} in a careful manner using
the Higgs portal DM models that are renormalizable, gauge invariant and
unitary~\cite{Baek:2011aa,Baek:2012uj,Farzan:2012hh,Baek:2012se}.
For high-energy collider studies, using an effective field theory (EFT)
could be dangerous, especially when we do not know the mass scales of
new degrees of freedom.
This is especially true for the dark matter physics, since nothing is
known about the DM mass, their interactions among themselves and with
the SM particles, as well as the masses of dark force mediators such as
a dark photon or the dark Higgs boson.

Earlier analyses of this issue are based on the following model
Lagrangians~\cite{Silveira:1985rk,Burgess:2000yq,Djouadi:2011aa,%
Djouadi:2012zc}:
\begin{eqnarray}
{\cal L}^{\rm EFT}_{\rm SSDM} &  = & \frac{1}{2} \partial_\mu S
 \partial^\mu S - \frac{1}{2} m_S^2 S^2 - \frac{\lambda_S}{4 !} S^4 
- \frac{\lambda_{HS}}{2} S^2 H^\dagger H
\\  \label{SFDM-EFT}
{\cal L}^{\rm EFT}_{\rm SFDM} & = & \overline{\chi} ( i
\slashed{\partial} - m_\chi ) \chi - 
\frac{\lambda_{\chi H}}{\Lambda} \overline{\chi} \chi H^\dagger H 
\\  \label{VDM-EFT}
{\cal L}^{\rm EFT}_{\rm  VDM} & = &  - \frac{1}{4} V_{\mu\nu} V^{\mu\nu}
+ \frac{1}{2} m_V^2 V_\mu V^\mu - \frac{\lambda_{VH} }{2} V_\mu V^\mu
H^\dagger H - \frac{\lambda_V}{4} ( V_\mu V^\mu )^2
\end{eqnarray}
The Lagrangian for scalar DM (1.1) is renormalizable and
unitary, and one can study scalar DM phenomenology without any
limitation.\footnote{There is an issue about Planck-scale suppressed
$Z_2$ breaking operator which would make EW scale scalar DM decay
fast.  We refer to Ref.~\cite{Baek:2014kna} for implementing global $Z_2$
to $Z_2$ subgroup of $U(1)$ dark gauge symmetry which resolves this
problem.
See also Ref.~\cite{Ko:2014nha} for comparison of global vs.\ local
$Z_3$ scalar DM models.}
On the other hand, the Lagrangians for singlet fermion DM (\ref{SFDM-EFT}) 
contains a dim-5 operator Higgs portal interaction, and eventually one has to 
consider its UV completions.
The simplest UV completion for the singlet fermion DM model with Higgs
portal have been constructed in Refs.~\cite{Baek:2011aa,Baek:2012uj} by
introducing a real singlet scalar mediator that couples to singlet
fermion DM and also to the SM Higgs fields in an $SU(2)$ singlet
combination, $H^\dagger H$.
Both DM phenomenology and vacuum structures of the model have been
studied in great detail.
After electroweak symmetry breaking, the singlet scalar mixes with the 
SM Higgs boson, which plays an important role in DM direct detections as
well as DM searches at colliders.

Likewise, the Higgs portal vector DM (VDM) model is problematic,
because it does not respect either unitarity or gauge invariance since
the VDM mass is given by hand.
Such drawbacks could be overcome in UV-complete VDM
models~\cite{Farzan:2012hh,Baek:2012se,Duch:2015jta,DiFranzo:2015nli}, 
where VDM becomes a dark gauge boson
associated with a local $U(1)_X$ dark gauge symmetry.  VDM gets massive
through a dark Higgs mechanism, and there would be a mixing between the
SM Higgs boson and the dark Higgs boson.
Then VDM becomes stable by {\it ad hoc} $Z_2$ symmetry 
plus charge conjugation symmetry under $U(1)_X$.  One can also construct 
models where VDM becomes absolutely stable due to the unbroken dark gauge 
symmetry, which has much richer structure in the dark sector, namely
dark monopole, massless dark photon and dark Higgs boson as well as
VDM~\cite{Baek:2013dwa}.

These models have a new degree of freedom, a singlet-like scalar boson,
which mixes with the SM Higgs boson.
Therefore at least two more parameters appear: the mass of the 2nd
scalar boson $m_{H_2}$ and the mixing angle $\alpha$, as well as the
coupling between the singlet scalar and DM.
One of the authors utilizes the 2nd scalar in order to explain the
galactic center $\gamma$-ray excess~\cite{Ko:2014gha,Ko:2014loa,%
Baek:2014kna,Ko:2015ioa} and to obtain a larger tensor-to-scalar ratio
in the Higgs portal assisted Higgs inflation scenario~\cite{Ko:2014eia}.
Also it was shown that the correlation between the upper bound on the
invisible Higgs decay branching ratio and the upper bound on the direct
detection cross section is modified in the renormalizable, unitary and
gauge invariant models~\cite{Baek:2014jga}.  Global analysis of the SM
Higgs signal strengths should include its possible mixing with a singlet
scalar in this kind of DM models~\cite{Chpoi:2013wga,
Cheung:2015dta,Cheung:2015cug}

The shortcomings of these effective Lagrangians for singlet fermion and
vector DM cases have been pointed out within the DM phenomenology~\cite{Baek:2011aa,Baek:2012uj}.
Especially the direct-detection cross-section (DM-nucleon scattering
cross-section) depends on the model Lagrangians, namely there is a generic
destructive interference between the SM Higgs boson and the second
singlet-like scalar boson.

One important question in the Higgs portal DM models is how to
distinguish three different cases at colliders and/or other
experiments.
In this paper, we study this issue at the ILC, postponing the same issue
at the LHC for future publication~\cite{lhc}. 
At the ILC, we can fix the initial beam energy and measure the 4-momenta
of the final $Z^0$ in the process $e^+ e^- \rightarrow Z^0$ + \ET, and
there are simple relations among $E_Z$, $M_{DD}$ and \ET:
\begin{eqnarray*}
M_{DD}^2 & = & s + m_Z^2 - 2 E_Z \sqrt{s} \ ,  \\
\ET & = & \frac{s - m_Z^2 + M_{DD}^2}{2\sqrt{s}} \ ,   \\
E_Z & = & \frac{s + m_Z^2 - M_{DD}^2}{2\sqrt{s}} \ ,   
\end{eqnarray*}
where $\sqrt{s}$ is  the total collision energy in the laboratory frame, $M_{DD}$ is the 
invariant mass squared of the DM pair ($D=S,\chi,V$ in the following sections).
Therefore one can reconstruct all the relevant kinematic variables
related with DM, $M_{DD}^2$ and $\ET$ at the ILC,
which renders us to study the Higgs portal  DM properties in clean ways. 

This paper is organized as follows. 
In Sec.~2, we describe the Higgs portal DM models for scalar, fermion
and vector DM.   
We set up the renormalizable, unitary and gauge invariant Lagrangians,
which can be used at an arbitrarily high-energy scale relevant to colliders, 
and often produce different results from the Higgs portal DM models with 
effective nonrenormalizable and gauge-variant interactions. 
First, we list the processes for the DM production at the ILC in the
Higgs portal DM models. 
Then, we present the detail analysis for the relevant process:
\begin{equation}
e^+(p_1) + e^-(p_2) \to h^*(q) + Z(p_Z) \to S(k_1) + S(k_2) + Z(p_Z) \ ,
\label{eq:process}
\end{equation}
for the scalar DM case, and the counter processes for the fermion and
vector DM models. 
In particular there are two scalar propagators contributing to this
process for the fermion and vector DM cases, as first pointed out in
Ref.~\cite{Baek:2015lna}. 
Then in Sec.~3, we give qualitative discussions on how to distinguish 3
different cases with the rate and shape analysis.
In Sec.~4, we describe the detailed analysis on the parameter
constraints at the ILC at $\sqrt{s}=500$ GeV, and compare our results
with those obtained from the Higgs portal DM models based on the
effective field theories.
Finally we conclude our analysis in Sec.~5.

\section{Model Lagrangians \label{sec:model}}

In this section, we define the simplified models for Higgs portal DM,
where DM can be either a scalar, fermion or vector particle.
It is important to start from model Lagrangians that are unitary and
renormalizable and invariant under full SM gauge symmetry.

At the ILC, Higgs portal DM can be produced through the following
processes~\cite{Matsumoto:2010bh,Kanemura:2011nm,Chacko:2013lna,%
Fedderke:2015txa,Andersen:2013rda}:
\begin{equation}
e^+ e^- \rightarrow \left\{ 
\begin{array}{cc} 
Z^0 H_{i=1,2} , &({\rm Higgs\mathchar`-strahlung}) ,  \\ 
\nu_e \overline{\nu_e} H_{i=1,2} , & (W^+ W^-  {\rm fusion}) ,  \\
e^+ e^-  H_{i=1,2} , & (Z^0 Z^0 {\rm fusion}) , 
\end{array}
\right.
\end{equation}
all of which are followed by the $H_{i=1,2}DD$ couplings.
Note that as we will see in the following, $H_1$ and $H_2$ are the
Higgs bosons portal to DM in the fermion and vector DM models.
There is no $H_2$ and $H_1 = H_{\rm SM}$ in the scalar DM case.
One can identify the 1st process by tagging $Z^0$ in the $\mu^+\mu^-$ or
$q\bar{q}$ channels.
The 2nd process is impossible to observe since there is no detectable
particle in the final-state.
The 3rd process has a unique event topology too, but the total
cross-section is more than 10 times smaller than the 1st process up to
$\sqrt{s}\simeq3$~TeV~\cite{Kanemura:2011nm}.
Therefore, the 1st process is the most promising process to observe at
the ILC with the current
proposal~\cite{Matsumoto:2010bh,Chacko:2013lna}.

In the following, we consider the 1st process as the signal of the
DM production in the Higgs portal DM models.
However, depending on the details of the models which satisfy the gauge
invariance, unitarity and renormalizability, as well as on the model 
parameter regions, the collider signatures can be different.
Thus, we have to perform the careful study on these detectability.

\subsection{Singlet scalar DM with Higgs portal}

In the scalar DM case, the model is very simple:
\begin{equation}
{\cal L}_{\rm SSDM} = \frac{1}{2} \partial_\mu S \partial^\mu S 
- \frac{1}{2} m_{0}^2 S^2 - \lambda_{HS} H^\dagger H S^2  
- \frac{\lambda_S}{4 !} S^4.
\end{equation}
The terms with odd numbers of $S$ is restricted by imposing $Z_2$
symmetry under which $S$ changes sign, but all the SM particles do not. 
Then $S$ will be stable and can make a good DM candidate.

 From this Lagrangian, we can calculate the amplitude for the process,
 (\ref{eq:process}):
\begin{align}
 i{\cal M}_{S} =& \bar v(p_2,\lambda_2)(-i\frac{g_Z}{2})
 \left[c_V^e\gamma^\mu-c_A^e\gamma^\mu\gamma_5\right] u(p_1,\lambda_1)
 \cdot
 \frac{-i(g_{\mu\nu}-\frac{P_\mu P_\nu}{m_Z^2})}{s-m_Z^2+im_Z\Gamma_Z}
 \nonumber \\
 &\cdot ig^{\nu\alpha}\frac{2m_Z^2}{v}\epsilon_\alpha(p_Z)
 \times \left[\frac{i}{t-m_h^2+im_h\Gamma_h}\cdot
 2i\lambda_{HS}v\right].
\end{align}
We define $P^\mu=p_1^\mu+p_2^\mu$, $c_V^e=-1/2+2s_W^2$ with
$s_W=\sin\theta_W$ where $\theta_W$ is the weak mixing angle, and
$c^e_A=-1/2$.
Then the amplitude ${\cal M}_S$ can be factorized into two parts: 
\begin{align}
 {\cal M}_{S} = {\cal
 M}_{h^*Z}\cdot\frac{2\lambda_{HS}v}{t-m_h^2+im_h\Gamma_h}.
\end{align}
The squared amplitude for the $h^*Z$ production part is 
\begin{align}
 \left|{\cal M}_{h^*Z}\right|^2 =
 {\cal P}(\lambda_1,\lambda_2)
\, \frac{8m_Z^4}{v^4}\left|r_Z(s)\right|^2
 \left(\frac{p_1\cdot p_Z p_2\cdot
 p_Z}{s^2}+\frac{m_Z^2}{s}\right),
\end{align}
where we define $P_{ee}(\lambda_1,\lambda_2) = (1-\lambda_1\lambda_2)
 \left(|c_V^e|^2+|c_A^e|^2\right)-2(\lambda_1-\lambda_2)\,{\rm
 Re}(c_V^e{c_A^{e*}})$ and $r_Z(s)=1/(1-m_Z^2/s+im_Z\Gamma_Z/s)$.
For the spin-averaged cross-section, ${\cal P}_{ee}\to\overline{\cal
 P}_{ee}=\left(|c_V^e|^2+|c_A^e|^2\right)$ is adopted.
The 3-body phase-space is given by
\begin{align}
 d\Phi_3(p_1+p_2;k_1,k_2,p_Z) &=
 \frac{dt}{2\pi}\cdot
 d\Phi_2(p_1+p_2;q,p_Z)\cdot d\Phi_2(q;k_1,k_2)\nonumber \\ &=
 \frac{dt}{2\pi}\cdot
 \frac{\hat\beta}{8\pi}\frac{d\hat\Omega}{4\pi}\cdot
 \frac{\beta_S}{8\pi}\frac{d\Omega_S}{4\pi},
\end{align}
where $t=q^2$, $\hat\beta=\lambda^{1/2}(1,m_Z^2/s,t/s)$ and
$\beta_D=\lambda^{1/2}(1,m_D^2/t,m_D^2/t)=\sqrt{1-4m_D^2/t}$ (for $D=S$,
$\chi$, $V$), with $\lambda(a,b,c)=a^2+b^2+c^2-2(ab+bc+ca)$.
The range of the kinematic variable $t$ is 
\[
4 m_D^2 \leq t \leq ( \sqrt{s} - m_Z )^2
\]
for a given $\sqrt{s}$ (the CM energy of the ILC).
$d\hat\Omega$ and $d\Omega_S$ are two-body phase-space volumes for the 
$h^*Z$ and the $SS$ systems, respectively.

Thus, the cross-section is straightforwardly calculated to be 
\begin{align}
 d\sigma_{S} &= C_S
 \frac{1}{2s}\left|{\cal M}_{S}\right|^2d\Phi_3 \nonumber \\
 &= \frac{dt}{2\pi}\cdot
 \frac{1}{2s}\left|{\cal M}_{h^*Z}\right|^2d\Phi_2(p_1+p_2;q,p_Z)
 \cdot C_S\frac{\beta_D}{8\pi}
 \left|\frac{2\lambda_{HS}v}{t-m_h^2+im_h\Gamma_h}\right|^2 .
\end{align}
Here $C_S$ is a symmetric factor, $C_S=1/2$, taking care of the
identical $S$'s in the final states.   By defining the total cross
section for $e^+e^-\to h^*Z$ as 
\begin{align}
 \sigma_{h^*Z}(s,t) &=
 \frac{1}{2s}\left|{\cal M}_{h^*Z}\right|^2d\Phi_2(p_1+p_2;q,p_Z)
 \nonumber\\
 &= {\cal P}_{ee}(\lambda_1,\lambda_2)
 \frac{1}{6s}\frac{m_Z^4}{v^4}\left|r_Z(s)\right|^2
 \frac{\hat\beta}{8\pi}\left[\hat\beta^2+\frac{12m_Z^2}{s}\right], 
\end{align}
and a form factor for the scalar DM production as
\begin{align}
 G_S(t) = C_S\frac{\beta_D}{8\pi}
 \left|\frac{2\lambda_{HS}v}{t-m_h^2+im_h\Gamma_h}\right|^2,
\end{align}
the $t$-distribution is given as
\begin{align}
 \frac{d\sigma_{S}}{dt} &= \frac{1}{2\pi}\sigma_{h^*Z}(s,t)\cdot
 G_S(t).
\end{align}
Note that, at lepton colliders, $t$ is observable from the
$Z$-boson momentum by $t=(p_1+p_2-p_Z)^2=s+m_Z^2-2\sqrt{s}E_Z$ where
$E_Z$ is the $Z$-boson energy in the C.M.\ frame of $e^+e^-$.
$\sigma_{h^*Z}$ depends on $t$ as well through $\hat\beta$.

\subsection{Model for singlet fermion DM with Higgs portal}

In the case of the Higgs portal fermion DM model, we assume that DM is a
singlet Dirac fermion $\chi$ with some nontrivial dark charge so that
it is distinguishable from right-handed neutrinos. 
Otherwise one can write down the Dirac neutrino Yukawa terms and $\chi$
would decay and cannot be a good cold DM candidate.
The simplest UV-completion of the Higgs portal fermion DM model can be
constructed by introducing a $SU(2)$-singlet scalar which has a vacuum
expectation value and a Yukawa interaction to DM giving its mass:
\begin{eqnarray}
{\cal L}_{\rm SFDM} &=& \overline{\chi} ( i \slashed{\partial} - m_\chi
 - y_\chi \phi ) \chi + \frac{1}{2} \partial_\mu \phi \partial^\mu \phi  -
 \frac{1}{2} m_0^2 \phi^2
\label{eq:portal}
\\
& -& \lambda_{H\phi} H^\dagger H \phi^2 - \mu_\phi \phi H^\dagger H -
 \mu_0^3 \phi
 - \frac{\mu_\phi}{3 !} \phi^3 - \frac{\lambda_\phi}{4 !} \phi^4.
\nonumber 
\end{eqnarray}
Expanding both fields around their VEVs
by $H\to\left(0,(v_H+h)/\sqrt{2}\right)^{\top}$ and $\phi\to
v_\phi+\phi$, we can derive the Lagrangian in terms of $h$ and $\phi$.
 After diagonalization of the mass matrix in the scalar sector, 
 \begin{align}
  \left(
  \begin{array}{c}
   h \\ \phi
  \end{array}
  \right) = \left(
\begin{array}{cc}
 \cos\alpha& \sin\alpha\\ -\sin\alpha & \cos\alpha
\end{array}
  \right)\left(
  \begin{array}{c}
   H_1 \\ H_2
  \end{array}
  \right) ,
 \end{align}
 DM $\chi$ couples with both $H_1$ and $H_2$.
The interaction Lagrangian of $H_1$ and $H_2$ with the SM fields and DM
$\chi$ is given by 
\begin{eqnarray}
 {\cal L}_{\rm int} & = &  - ( H_1 \cos\alpha + H_2 \sin\alpha )
  \left[ \sum_f \frac{m_f}{v_H} \overline{f} f  - \frac{2 m_W^2}{v_H}
   W_\mu^+ W^{-\mu} - \frac{m_Z^2}{ v_H} Z_\mu Z^\mu \right]
  \nonumber \\ 
 & + & y_\chi ( H_1 \sin\alpha - H_2 \cos\alpha ) \overline{\chi}\chi \ ,
 \end{eqnarray}
following the convention of Ref.~\cite{Baek:2011aa}.
We identify the observed 125~GeV scalar boson as $H_1$.
The mixing between $h$ and $\phi$ leads to a universal suppression
factor of the Higgs signal strengths at the LHC, independent of
production and decay channels~\cite{Baek:2011aa}.
From the current data on Higgs signal strengths and the upper bound on
the Higgs invisible branching ratio, one can derive an upper bound,
$\sin\alpha\lesssim0.53$~\cite{Khachatryan:2014jba,Aad:2015gba,Couplings}.
This bound is still quite weak and should be improved in the future
experiments.

Defining $\lambda_\chi=y_\chi\sin\alpha\cos\alpha$, the scattering
amplitude for the process,
\begin{align}
e^+(p_1) + e^-(p_2) \to H_1/H_2(q) + Z(p_Z) \to \chi(k_1) +
 \bar\chi(k_2) + Z(p_Z) \ ,
\end{align}
is written as 
\begin{align}
 {\cal M}_{\chi} = {\cal M}_{h^*Z}\cdot\lambda_\chi
 \left[\frac{1}{t-m_{H_1}^2+im_{H_1}\Gamma_{H_1}}
 - \frac{1}{t-m_{H_2}^2+im_{H_2}\Gamma_{H_2}}\right]\bar{u}(k_1)v(k_2).
\end{align}
Thus, the squared matrix elements are
\begin{align}
 \sum\left|{\cal M}_{\chi}\right|^2=\left|{\cal M}_{h^*Z}\right|^2
 \lambda_\chi^2
 \left|\frac{1}{t-m_{H_1}^2+im_{H_1}\Gamma_{H_1}}
 - \frac{1}{t-m_{H_2}^2+im_{H_2}\Gamma_{H_2}}\right|^2
 \sum\left|\bar{u}v\right|^2, 
\end{align}
where the spin-sum of the fermion DM wave-functions is evaluated to be
\begin{align}
 \sum\left|\bar{u}v\right|^2 = 2(t-4m_\chi^2) = 2t\beta_\chi^2.
\end{align}
Thus, we obtain 
\begin{align}
 \frac{d\sigma_{\chi}}{dt} &= \frac{1}{2\pi}\sigma_{h^*Z}(s,t)\cdot
 G_\chi(t),
\end{align}
where
\begin{align}
 G_\chi(t) = C_\chi\frac{\beta_\chi^3}{8\pi}\cdot 2\lambda_\chi^2 t\cdot
 \left|\frac{1}{t-m_{H_1}^2+im_{H_1}\Gamma_{H_1}}
 - \frac{1}{t-m_{H_2}^2+im_{H_2}\Gamma_{H_2}}\right|^2.
\end{align}
The symmetric factor for this case is $C_\chi=1$.

\subsection{Gauge invariant unitary model for vector DM with Higgs portal }

There are a number of different models for stable or long-lived vector
DM $V_\mu$ with Higgs portal.
The simplest model would be a phenomenological model where a discrete
$Z_2$ symmetry ($V_\mu \rightarrow - V_\mu$) is imposed by
hand~\cite{Farzan:2012hh,Baek:2012se,Duch:2015jta}.
In order to construct a renormalizable and unitary model, it is
important to assume a dark gauge symmetry $U(1)_X$ and dark Higgs
$\Phi$ whose VEV provides a nonzero mass to vector DM $V_\mu$ and
produces a dark Higgs $\varphi$ as a remnant of the Higgs mechanism:
\begin{equation}
{\cal L} = -\frac{1}{4} V_{\mu\nu} V^{\mu\nu} + D_\mu \Phi^\dagger D^\mu \Phi 
- \frac{\lambda_\Phi}{4} ( \Phi^\dagger \Phi - \frac{v_\phi^2}{2} )^2 
- \lambda_{H\Phi}   ( H^\dagger H - \frac{v^2}{2} ) 
( \Phi^\dagger \Phi - \frac{v_\phi^2}{2} ).
\end{equation}

One can also consider more sophisticated models where the aforementioned 
{\it ad hoc} $Z_2$ symmetry is implemented to some local dark gauge symmetries.
There are basically two different categories in this class.  
\begin{itemize}
\item VDM is stable due to unbroken gauge symmetry~\cite{Baek:2013dwa}: 
One of the present authors 
constructed a hidden sector monopole model where the renowned 't~Hooft-Polyakov 
monopole is put in the hidden sector, which is connected to the SM sector through 
the Higgs portal interaction. There are two stable DM in this case: hidden monopole which 
is stable due to topological reason, and vector DM which is stable due to unbroken 
$U(1)_X$ subgroup. There is massless dark photon associated with unbroken $U(1)_X$,
and it can generate strong self-interaction between dark matters, and would contribute
to the dark radiation at the level of $\sim 0.1$ which is perfectly consistent with Planck data.
\item VDM is stable at renormalizable level, but could decay and is long lived 
when we consider higher dimensional operators~\cite{Hambye:2008bq}. 
This happens if the dark gauge group $SU(2)_X$ is completely broken by $SU(2)_X$ doublet complex dark Higgs, for example.
In this case the dark gauge symmetry is completely broken, and the massive VDM is 
not stable due to the presence of higher dimensional nonrenormalizable operators.
\end{itemize}

In this paper, we do not consider these sophisticated models, and will consider 
phenomenological VDM models, which could be considered as a simplified
VDM model:
\begin{equation}
{\cal L}_{\rm VDM} = - \frac{1}{4} V_{\mu\nu} V^{\mu\nu} 
+ \frac{1}{2} m_V^2 V_\mu V^\mu (1 + \frac{\varphi}{v_\varphi} )^2 
+ \frac{1}{2} \partial_\mu \varphi \partial^\mu \varphi -
\lambda_{H \varphi} ( v h + \frac{1}{2}  h^2 ) ( v_\varphi \varphi +
\frac{1}{2} \varphi^2).
\end{equation}
Similarly to the fermion DM model, $h$ and $\varphi$ are expressed in
terms of the mass eigenstates $H_1$ and $H_2$ with mixing angle
$\alpha$.
For the purpose of studying the collider signatures, it would be
enough to consider the following simplified VDM with Higgs portal as, 
ignoring the triple and quartic couplings of $H_1$ and
$H_2$:\footnote{Higgs pair productions will be discussed in the future 
publication.}
\begin{eqnarray}
{\cal L}_{\rm VDM} & = & - \frac{1}{4} V_{\mu\nu} V^{\mu\nu} 
+ \frac{1}{2} m_V^2 V_\mu V^\mu 
+ \frac{1}{2} \partial_\mu \varphi \partial^\mu \varphi 
- \frac{1}{2}\frac{2m_V^2}{ v_\varphi} V_\mu V^\mu ( H_1 \sin\alpha -
H_2 \cos \alpha )
\nonumber  \\
& - &  ( H_1 \cos\alpha + H_2 \sin\alpha ) \left[ \sum_f \frac{m_f}{v_H} 
 \bar{f} f  - \frac{2 m_W^2}{v_H} W_\mu^+ W^{-\mu} - \frac{m_Z^2}{ v_H}
 Z_\mu Z^\mu \right] .
\end{eqnarray}

By defining $g_V=2m_V/v_\varphi$ and
$\lambda_V=g_V\cos\alpha\sin\alpha$, the scattering amplitude for the
process 
\begin{align}
e^+(p_1) + e^-(p_2) \to H_1/H_2(q) + Z(p_Z) \to V(k_1) +
 V(k_2) + Z(p_Z) \ ,
\end{align}
is given as
\begin{align}
 {\cal M}_{V} = {\cal M}_{h^*Z}\cdot\lambda_Vm_V
 \left[\frac{1}{t-m_{H_1}^2+im_{H_1}\Gamma_{H_1}}
 - \frac{1}{t-m_{H_2}^2+im_{H_2}\Gamma_{H_2}}\right]
 \epsilon_1^*(k_1)\cdot\epsilon_2^*(k_2).
\end{align}
The squared amplitude is evaluated to be
\begin{align}
 \sum\left|{\cal M}_{V}\right|^2=\left|{\cal M}_{h^*Z}\right|^2
 \cdot(\lambda_Vm_V)^2
 \left|\frac{1}{t-m_{H_1}^2+im_{H_1}\Gamma_{H_1}}
 - \frac{1}{t-m_{H_2}^2+im_{H_2}\Gamma_{H_2}}\right|^2
 \sum\left|\epsilon_1^*\cdot\epsilon_2^*\right|^2,
\end{align}
where the spin-sum of the vector DM wave-functions is calculated as
\begin{align}
 \sum\left|\epsilon_1^*\cdot\epsilon_2^*\right|^2 =
 2+\frac{(t-2m_V^2)^2}{4m_V^4} = \frac{t^2}{4m_V^4}
 \left(1-\frac{4m_V^2}{t}+\frac{12m_V^4}{t^2}\right).
\end{align}
Thus, the $t$ distribution is obtained as
\begin{align}
 \frac{d\sigma_{V}}{dt} &= \frac{1}{2\pi}\sigma_{h^*Z}(s,t)\cdot
 G_V(t),
\end{align}
where
\begin{align}
G_V(t) = C_V
 \frac{\beta_V}{8\pi}\cdot\frac{\lambda_V^2t^2}{4m_V^2}
 \left(1-\frac{4m_V^2}{t}+\frac{12m_V^4}{t^2}\right)\cdot
 \left|\frac{1}{t-m_{H_1}^2+im_{H_1}\Gamma_{H_1}}
 - \frac{1}{t-m_{H_2}^2+im_{H_2}\Gamma_{H_2}}\right|^2.
\end{align}
The symmetric factor is $C_V=1/2$.

\subsection{Comparison of three models}

Before we proceed further, let us make comments on three Higgs portal DM
models for scalar, fermion and vector DMs, (2.2), (2.11) and (2.20),
respectively.
In all the cases, we have imposed $Z_2$ symmetry which stabilize DM,
$S$, $\chi$ and $V_\mu$. 
Note that the scalar sectors of these three models are not symmetric:
there is only one mediator~($H$) in the scalar DM case in Eq.~(2.2),
whereas there are two mediators both in the fermion DM case $(H, \phi)$
in Eq.~(2.11), and in the vector DM cases, $(H, \Phi)$ in Eq.~(2.20).
This is because of the gauge invariance and renormalizability. 
Singlet fermion DM $\psi$ cannot have renormalizable couplings to
the SM Higgs field, and one has to introduce a singlet scalar that can
couple to $\overline{\chi} \chi$ and mix with the SM Higgs field by
$\phi H^\dagger H$ and $\phi^2 H^\dagger H$ terms, as in Eq.~(2.11).
Likewise, the vector DM mass cannot be given by hand as in Eq.~(1.3). 
It has to be generated, for example, by dark Higgs mechanism by nonzero
VEV of $\Phi$ in Eq.~(2.20).
On the other hand, this is not the case for scalar DM, since we can
have a gauge invariant and renormalizable $S^2 H^\dagger H$ operator as
in Eq.~(2.2). 

In case of fermion or vector DM, these two scalar mediators always
appear in combination of Eq.~(2.15) or (2.24):\footnote{%
We assume that $H_1$ is the 125~GeV scalar boson observed at the LHC. }
\begin{equation}
 \frac{1}{t - m_{H_1}^2 + i m_{H_1} \Gamma_{H_1}} - 
  \frac{1}{t - m_{H_2}^2 + i m_{H_2} \Gamma_{H_2}} \ ,
\end{equation}
whereas for the scalar DM case only the SM Higgs plays the role of
mediator, 
\begin{equation}
\frac{1}{t - m_{H_1}^2 + i m_{H_1} \Gamma_{H_1}}  \ .
\end{equation}
Note that the relative size between two propagators, $-1$,  in
Eq.~(2.29) is originated from the $SO(2)$ nature of the rotation matrix
from the interaction eigenstates to the mass eigenstates in Eq.~(2.12).

It may be possible to make three models more symmetric if we introduce
additional new fields.
For example, we can introduce one more singlet scalar $\phi$ in the
scalar DM case with the following additional Lagrangian:
\begin{eqnarray}
\Delta {\cal L}_{\rm SSDM} & =  & \frac{1}{2} \partial_\mu \phi
 \partial^\mu \phi - \frac{1}{2} m_\phi^2 \phi^2 - \mu_\phi^{' 3} \phi -
 \frac{\lambda_\phi}{4} \phi^4 \\
 & - &  \mu_{\phi H} \phi H^\dagger H - \frac{1}{2} \mu_{\phi S} \phi S^2 
  - \frac{1}{2} \lambda_{\phi H} \phi^2 H^\dagger H - \frac{1}{4} \phi^2
  S^2 . 
\nonumber  
\end{eqnarray}
Then there will be two scalar mediators, $H_1$ and $H_2$, but the
relative sign and magnitudes of these two contributions to the processes
we consider would be completely free, and is not fixed to be $-1$ as in
Eq.~(2.31).
This is because both the singlet scalar $\phi$ and the SM Higgs $H$ can
have renormalizable couplings to scalar DM $S$. 
We do not consider this case further, since it is not minimal in terms of
the number of degrees of freedom. 

This difference in the number of force mediators in the scalar DM and in
the fermion/vector DM cases will generate the difference in the various
differential distributions studied (see Fig.~1, for example).
And this difference will make the high-$t$ behaviors of the amplitudes
very different, see Eqs.~(3.1)-(3.5) and the discussions.

\section{How to distinguish 3 cases at the ILC ?}

As we have evaluated in the previous section, the $t~(=M_{DD}^2)$
spectrum is given by
\[
\frac{d\sigma}{dt} \propto F(s,t) \times G_D(t)
\]
where the $t$-dependent form factor $G_D(t)$ is given by
\begin{eqnarray}
{\rm SSDM:}\quad  G_S(t) &\sim& \beta_D\frac{1}{(t-m_H^2)^2 + m_H^2
 \Gamma_H^2} ,
\\
{\rm SFDM:}\quad G_\chi(t) &\sim&
\beta_D^3\left| \frac{1}{t - m_{H_1}^2 + i m_{H_1}\Gamma_{H_1} }
- \frac{1}{t - m_{H_2}^2 + i m_{H_2} \Gamma_{H_2}} \right|^2
~\left( t - 4 m_\chi^2\right)
\nonumber \\
& \rightarrow & | \frac{1}{t^2} |^2 \times t \sim \frac{1}{t^3}
 ~({\rm as}~ t \rightarrow \infty\,{\rm
 for\,fixed}\,m_{H_1}\,{\rm and}\,m_{H_2}) , \\
{\rm VDM:}\quad G_V(t) &\sim& 
\beta_D\left|  \frac{1}{t - m_{H_1}^2 + i m_{H_1} \Gamma_{H_1} } 
- \frac{1}{t - m_{H_2}^2 + i m_{H_2} \Gamma_{H_2}} \right|^2~\left[ 2 +
\frac{(t-2m_V^2)^2}{4 m_V^4} \right]
\nonumber \\
& \rightarrow &  | \frac{1}{t^2} |^2 \times t^2 \sim  \frac{1}{t^2}
 ~({\rm as} ~t \rightarrow \infty\,{\rm
 for\,fixed}\,m_{H_1}\,{\rm and}\,m_{H_2}). 
\end{eqnarray}
The signal distribution arises at $t=(2m_D)^2$, thus by measuring the
threshold of the signal excess, the DM mass can be directly determined.
In addition, the threshold slope of the signal excess depends on the
spin of DM; for scalar and vector DM models it behaves as
$\propto\beta_D$, while for the fermion DM model it behave as
$\beta_D^3$.
This is also useful to distinguish the spin of DM by a kinematical
method~\cite{Asano:2011aj}.

The $t$ distribution in the high-$t$ region is sensitive to the
unitarity of the models.
If we ignore the 2nd Higgs propagator and identify $m_{H_1} = m_H$ 
(the discovered Higgs boson), we would have 
\begin{eqnarray}
 {\rm SFDM:}\quad
  G_\chi(t) &\sim& \beta_D^3 \frac{1}{(t-m_H^2)^2 + m_H^2\Gamma_H^2}
 ~\left( t - 4 m_\chi^2 \right) 
\nonumber \\
& \rightarrow & \frac{1}{t} ~({\rm as} ~t \rightarrow \infty) ,
\\
 {\rm VDM:}\quad
  G_V(t) &\sim& \beta_D\frac{1}{(t-m_H^2)^2 + m_H^2\Gamma_H^2}
 ~\left[ 2 + \frac{(t-2m_V^2)^2}{4 m_V^4} \right]
\nonumber \\
& \rightarrow &  {\rm constant} ~({\rm as} ~t \rightarrow \infty).
\end{eqnarray}
These results indicate the violation of unitarity in the total cross
section at large $s$ from large-$t$ region, and the EFT descriptions
based on (1.2) and (1.3)  would  become unreliable.
Note that ignoring the propagator of the 2nd Higgs would be justified if
$m_{H_2} \gg \sqrt{s}$.
On the other hand, in the UV-completed approach, the distribution
behaves well convergent at high-$t$.

Therefore, one would be able to determine the type of DM by observing
 the shape of the signal distribution.
Having enough number of bins and data, we can test by
 $\chi^2$-minimization to determine whether the observed \ET
 distribution follows that of scalar, fermion or vector DM with Higgs
 portal. 
Note that this procedure is possible at the ILC, and not at LHC, since
at the ILC the CM energy $\sqrt{s}$ is fixed so that one can factor out
the phase-space factor.
On the other hand, at hadron colliders, the parton-level CM energy
$\sqrt{\hat{s}}$ is not fixed so that we cannot factor out the
 phase-space factor in an unambiguous manner.
Note that for scalar DM, $G_S(t)$ is completely fixed by the SM
Higgs propagator, and there is no free parameter except $m_D$.
Therefore it would be straightforward to check if the observed \ET
distribution can be fitted by the SM Higgs propagator or not.
For the SFDM or VDM models, the fitting would be more complicated, since
in this case, there are 4 parameters: namely,
\[
\sin\alpha, ~~m_{H_2}, ~~ \Gamma_{H_2} , ~~m_{\rm D}.
\]
Note that we have to regard $\Gamma_{H_2}$ and $\sin\alpha$
independently, since $H_2 \rightarrow H_1 H_1$ can be newly open, which
calls a new parameter that could be traded with $\Gamma_{H_2}$.
With these 4 parameters, we can fit the \ET spectrum or the recoil-mass
spectrum, $M_{\rm rec}=\sqrt{t}$, and determine
the physical parameters in the SFDM or VDM models.

\begin{figure}[ht]
 \includegraphics[width=0.5\textwidth]{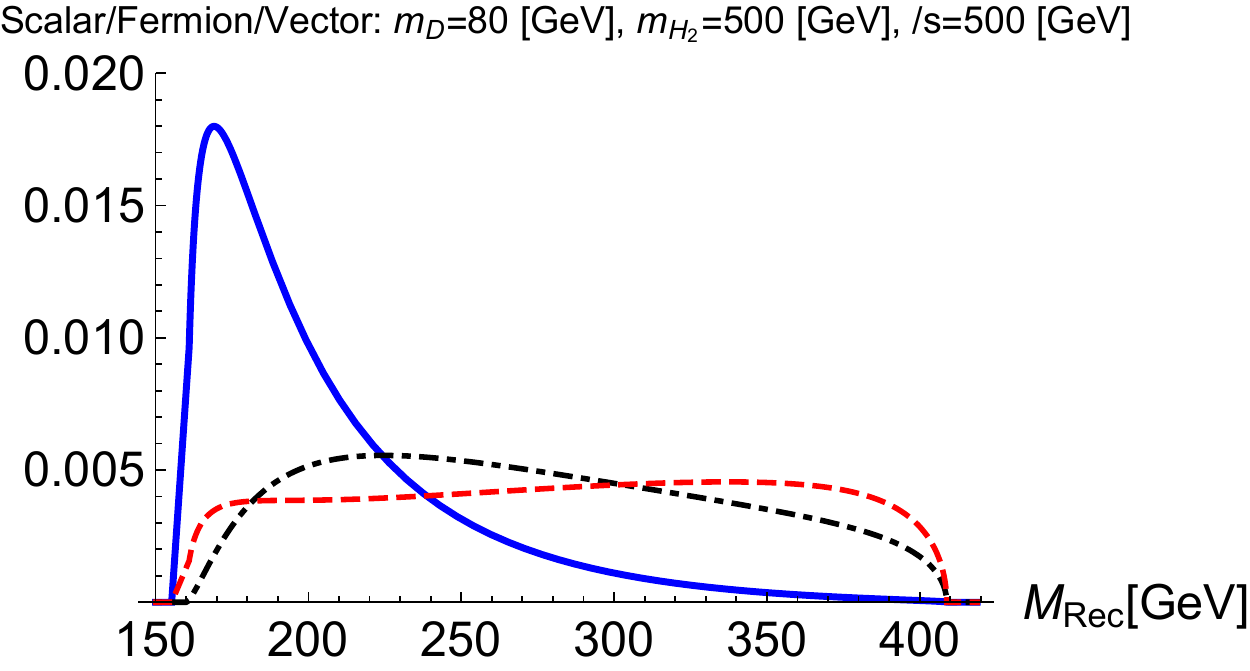}
 \includegraphics[width=0.5\textwidth]{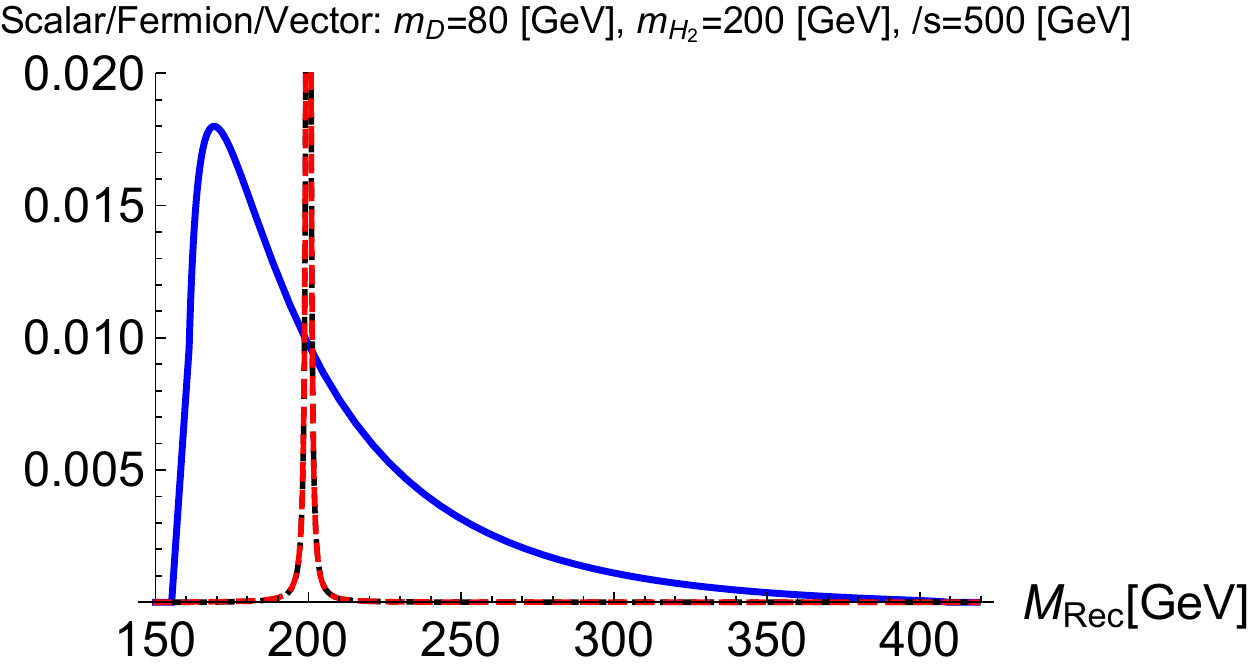}
 \\[5mm]
 \includegraphics[width=0.5\textwidth]{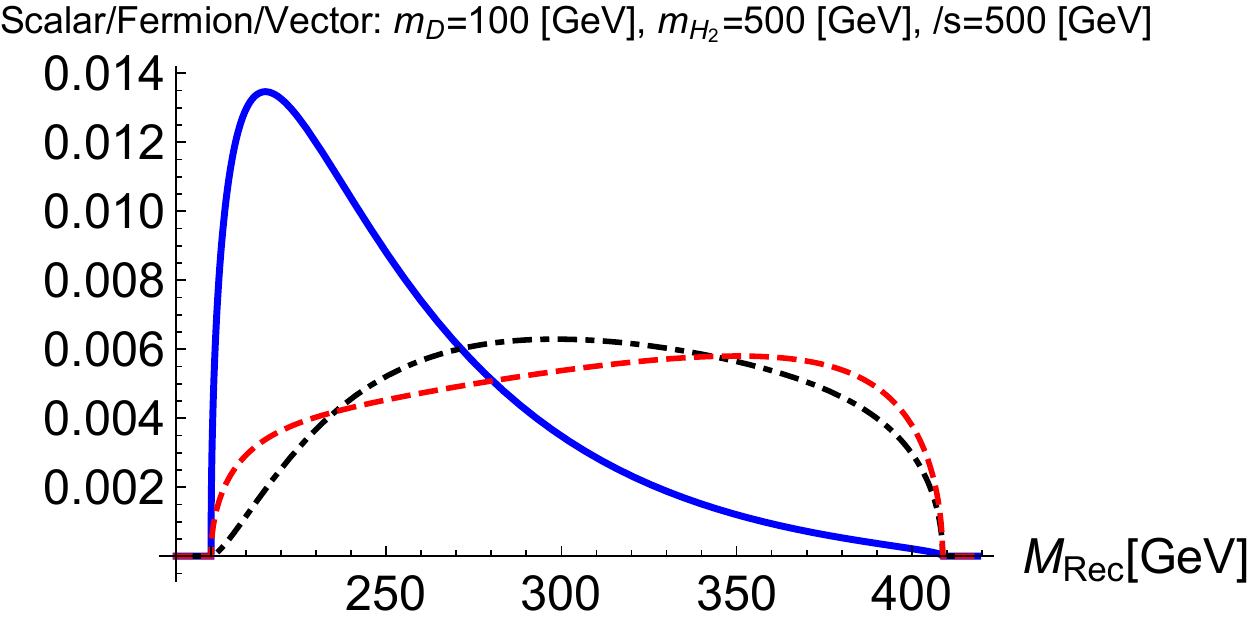}
 \\[5mm]
 \includegraphics[width=0.5\textwidth]{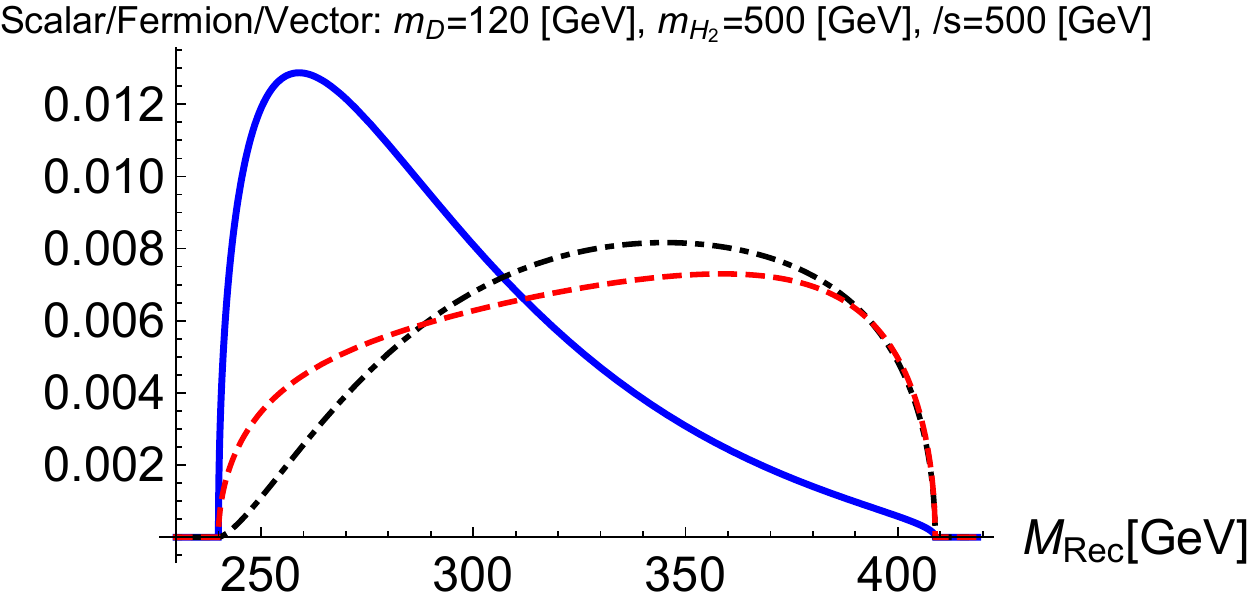}
 \includegraphics[width=0.5\textwidth]{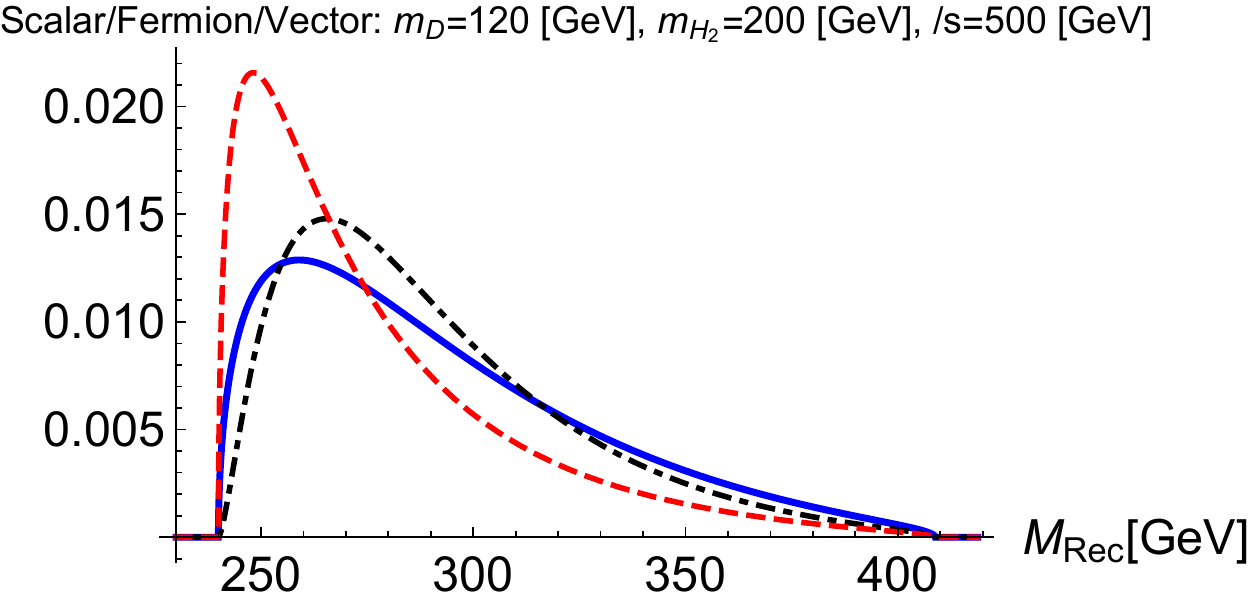}
 \caption{Normalized recoil-mass distribution in $e^+e^-\to
 ZH_1(/H_2^{(*)})\to ZDD$ at $\sqrt{s}=500$~GeV. Blue: scalar DM,
 Black: fermion DM, Red: vector DM.}\label{fig:1}
\end{figure}

In Fig.~\ref{fig:1}, the normalized recoil-mass distribution
$1/\sigma\cdot d\sigma/dM_{\rm rec}$ is evaluated for the scalar,
fermion and vector DM models for various sets of ($m_D$, $m_{H_2}$) at
$\sqrt{s}=500$~GeV.
$\Gamma_{H_2}=0.1$~GeV is used for simplicity.
The characteristic threshold behavior as well as the large
recoil-mass tail can be understood by the analytically-calculated form
factors given in the previous section.
For $2m_D\le m_{H_2}\le \sqrt{s}-m_Z$, on-shell $H_2$ can be produced and
subsequently decay into a pair of the DM particle.
In such cases, the recoil mass distribution shows a sharp peak at
$M_{\rm rec}=m_{H_2}$, and no difference can be observed between the
fermion DM and vector DM models. 
In addition, if $m_D\le m_{H_1}/2$, the recoil mass distribution shows
another peak at $M_{\rm rec}=m_{H_1}$ whose strength is expected to be
smaller than the peak at $m_{H_2}$ because of the constraints so far
[see Sec.~4.1].
Spin discrimination is still difficult since it behaves as a sharp peak
for each DM model.
For $2m_D\ge m_{H_2}$ or $m_{H_2}\ge \sqrt{s}-m_Z$, no sharp peak can be
observed, because on-shell $H_2$ cannot be produced because of the
limited collision energy or because the on-shell $H_2$ cannot decay into
a DM pair, so that the DM pair is produced through the off-shell $H_1$ and
$H_2$.
The recoil-mass distributions then behave as smooth curves depending on
the DM model as well as the masses of DM and $H_2$.
Thus, by measuring the shape of the distribution, one can determine the
type of DM and its mass, as well as the mass of the second Higgs
portal to DM.

%

\section{Parameter Constraints}

\begin{figure}[th]
 \begin{center}
  \includegraphics[width=0.5\textwidth]{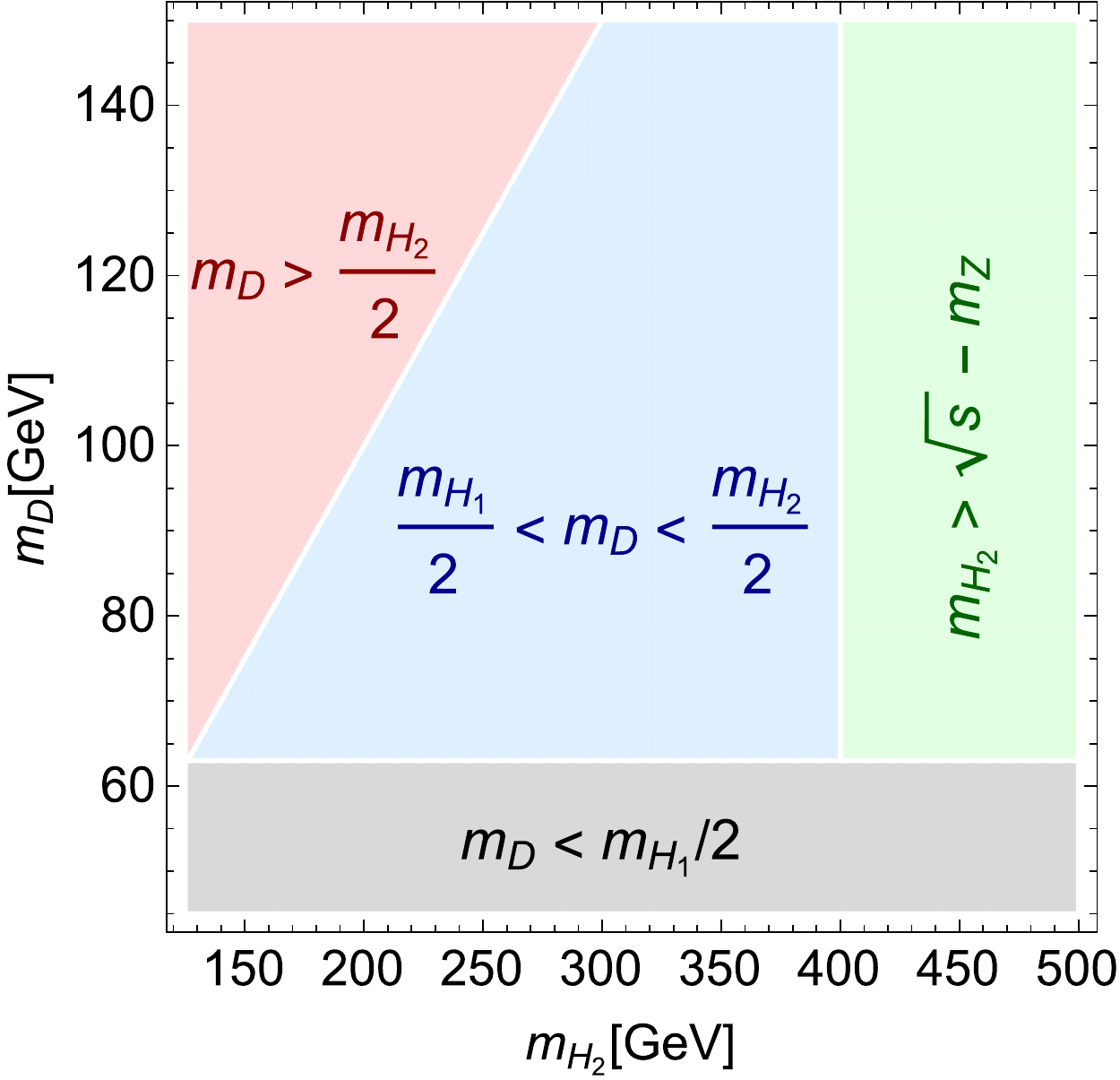}
  \caption{Dividing parameter space in the $(m_{H_2}$-$m_D)$
  plane by the plausible collider signature in the fermion and vector 
  DM models.
  In the scalar DM model, $H_2$ is absent.}\label{fig:mH2mD}
 \end{center}
\end{figure}

In this section, we discuss searches for the Higgs portal DM models at
the future lepton colliders. 
Depending on the masses of DM and the second Higgs boson in the
fermion and vector DM models, the search strategy at colliders can be
different. 
In Fig.~\ref{fig:mH2mD}, we divide the parameter space in
the $(m_{H_2}$-$m_D)$ plane in terms of the plausible collider signature
to search for in each parameter region in the fermion and vector DM
models.
In the scalar DM model, because of the absence of $H_2$, the parameter
region can be simply divided by $m_D<m_{H_1}/2$ or $m_D>m_{H_1}/2$, namely
whether the observed Higgs boson can decay into the DM pair or not. 
In the following subsections, we discuss collider signals of DM
production in each region of the parameter space, then further 
discuss the method to distinguish models, and to determine the model
parameters.

\subsection{$m_D\le m_{H_1}/2$}
In the case of $m_D\le m_{H_1}/2$, a DM pair can be produced in the
decay of $H_1$.
This can be seen as an invisible decay of the Higgs boson.
The limit on the branching ratio of the invisible decay has been
obtained as ${\cal B}(h\to{\rm inv.})\lesssim 0.25$ at the LHC~\cite{%
Aad:2015txa,Aad:2015pla,CMS:2015dia,CMS:2015naa}.
We see how the model parameters can be constrained by this measurement.
In the scalar DM model, the partial decay width for $H_1\to SS$ is
proportional to $\lambda_{HS}^2$.
By denoting $\Gamma(H_1\to SS)=\lambda_{HS}^2\Gamma_{0}$, the
experimental constraint of ${\cal B}(h\to {\rm inv}.)<{\cal B}(h\to {\rm
inv}.)|_{\rm exp.}\equiv X$, we obtain the limit on $\lambda_{\rm HS}$
as
\begin{align}
 \lambda_{HS}^2 < \frac{X}{1-X}
 \frac{\Gamma^{\rm SM}_{h}}{\Gamma_0}. 
\end{align}

For the fermion and vector DM models, the partial decay width for
$H_1\to \chi\bar{\chi}$ ($VV$) is proportional to $y_\chi^2s_\alpha^2$
($\lambda^2_V s_\alpha^2$).
On the other hand, the partial widths for the decay into SM particles
are all suppressed by $c^2_\alpha$.
By writing $\Gamma(H_1\to F\bar F)=y_\chi^2s_\alpha^2\Gamma_D$
[$\Gamma(H_1\to F\bar F)=g_V^2s_\alpha^2\Gamma_D$], 
the constraint on the branching ratio ${\cal B}(h\to {\rm inv}.)<X$
gives
\begin{align}
y_\chi^2 [g_V^2]\cdot t_\alpha^2<\frac{X}{1-X}
\frac{\Gamma_{\rm SM}}{\Gamma_D}.
\end{align}
Here, $t_\alpha=\tan\alpha$.
We evaluate the current limit by the LHC Run-I measurement, ${\cal
B}(h_{\rm SM}\to {\rm inv}.)\le0.25$~\cite{Aad:2015pla}, and also 
the accessible limits at future experiments, ${\cal B}(h_{\rm SM}\to
{\rm inv}.)\le0.0065$ at the ILC 500~GeV with 
500~fb$^{-1}$, and $\le0.0032$ with 1600~fb$^{-1}$~\cite{Fujii:2015jha}.
The limits are obtained on $\lambda_{HS}$, $y_\chi t_\alpha$ and
$g_Vt_\alpha$  in the scalar, fermion and vector DM models, respectively.
For the fermion (vector) DM model, the constraint on the coupling
$y_\chi$ ($g_V$) becomes weak for small $s_\alpha$.
The upper bound on $s_\alpha$ has been obtained by measuring
the signal strength of Higgs-gauge-gauge coupling $\kappa_V$ at the LHC
Run-I, which is equal to $c_\alpha$ in our models.
Current limit is about $\kappa_V\gtrsim 0.85$ at the 68\% C.L.~\cite{%
Khachatryan:2014jba,Aad:2015gba,Couplings}, thus
$s_\alpha\lesssim0.53$.
For the reference, in the cases of $c_\alpha=0.9$, 0.95 and 0.99,
$t_\alpha$ are $\simeq0.48$, 0.33 and 0.14, respectively.

\begin{figure}[ht]
 \includegraphics[width=0.5\textwidth]{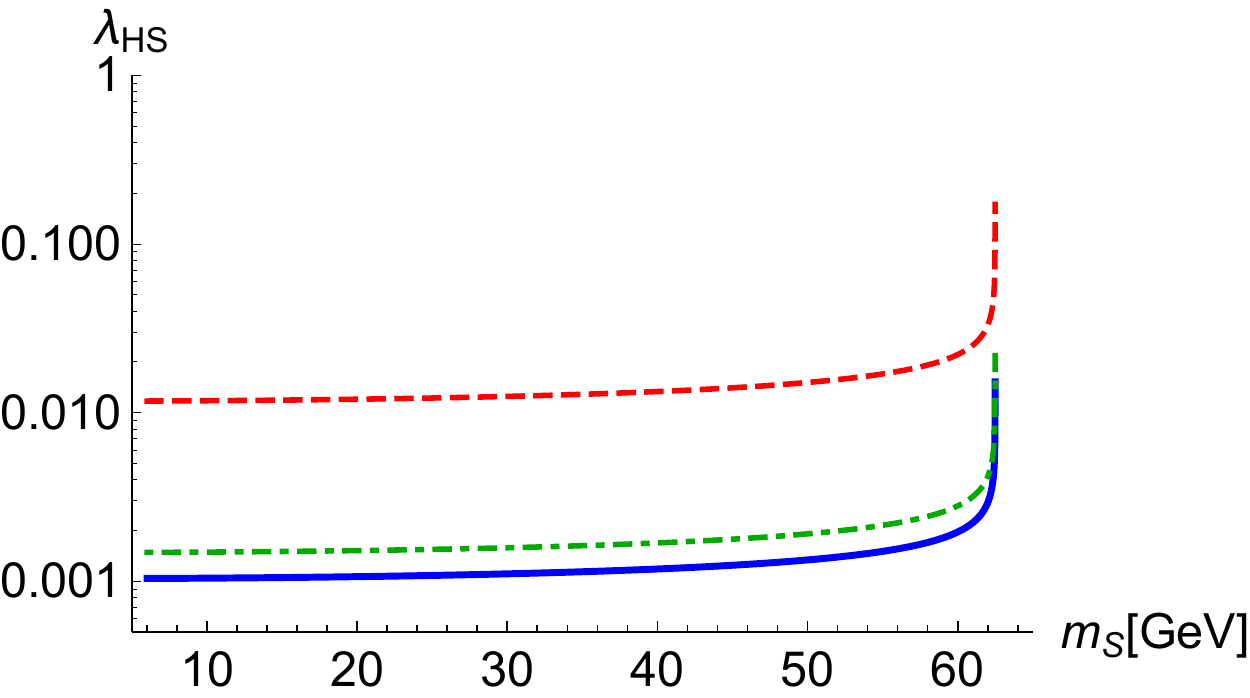}
 \quad
 \includegraphics[width=0.5\textwidth]{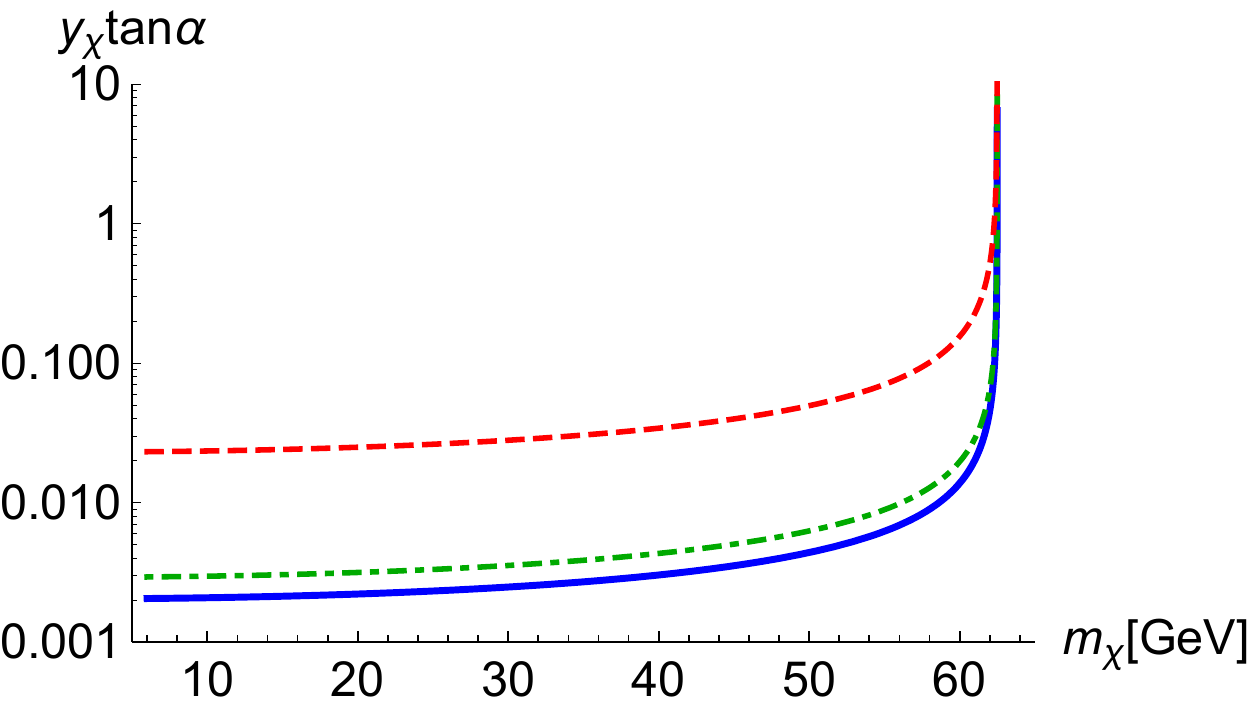}
 \quad
 \includegraphics[width=0.5\textwidth]{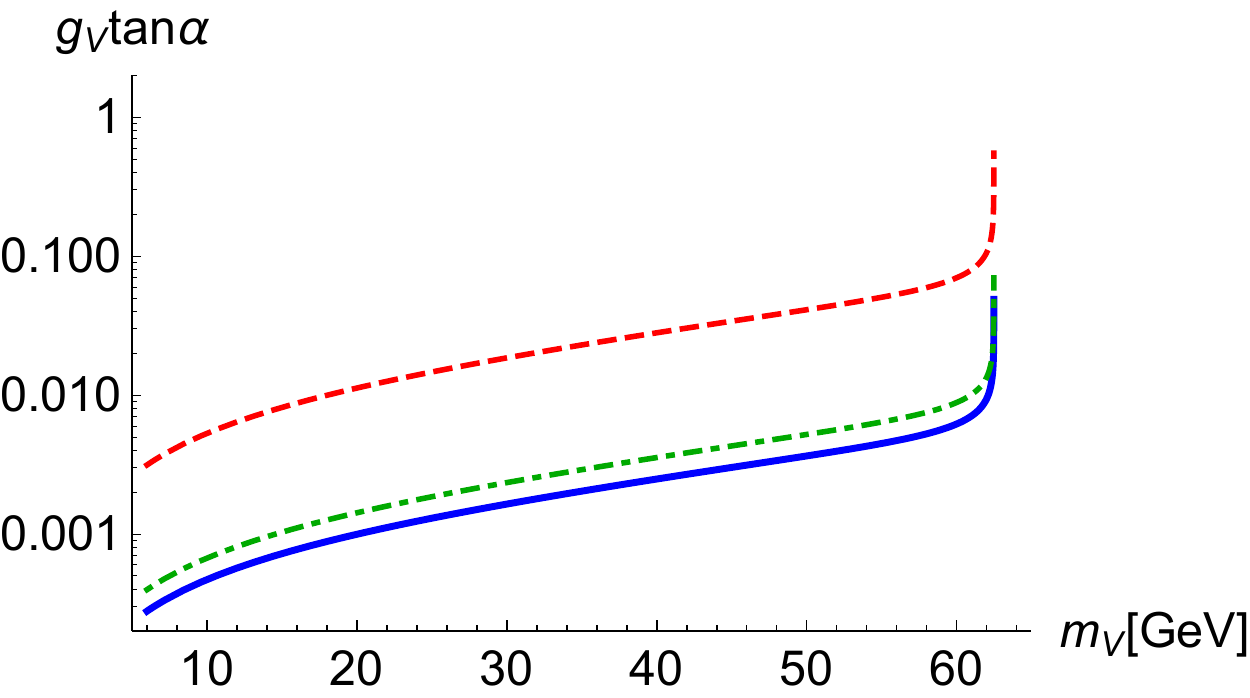}
 \caption{Current and future limits on the parameters in the scalar,
 fermion and vector DM models with $m_D\le m_{H_1}/2$.
 Constraints on the branching ratio of the Higgs invisible decay at the
 LHC Run-I (red, dashed), ILC 500~GeV with 500~fb$^{-1}$ (green
 dot-dashed), and 1600~fb$^{-1}$ (blue solid) are
 considered.}\label{fig:HiggsInvisible}
\end{figure}

In the top-left, top-right and bottom panels in
Fig.~\ref{fig:HiggsInvisible}, current and future limits on the
parameters in the scalar, fermion and vector DM models with $m_D\le
m_{H_1}/2$ are plotted as a function of $m_D$, respectively.
In the scalar DM model, $\lambda_{HS}$ is constrained to be
$\gtrsim0.01$ for $m_D\lesssim m_{H_1}/2$.
At the future ILC measurements, $\lambda_{HS}\gtrsim0.001$ will be
explored. 
In the fermion and vector DM models, the constraints on $y_\chi t_\alpha$
($g_V t_\alpha$) are $\gtrsim0.01$ by current LHC measurement, and
will be $\gtrsim0.001$ by future ILC measurements.
By observing non-zero $s_\alpha$ in future measurements, limits on the
Higgs-DM-DM coupling can be derived. 

\subsection{$m_D\ge m_{H_1}/2$}

In the scalar DM model with $m_D\ge m_{H_1}/2$, a DM pair is
produced through the off-shell $H_1$.
The collider signal for this case can be an excess in events with
a $Z$-boson plus missing energy with a large recoil-mass, $M_{\rm
rec.}\ge 2m_D$.
We consider muonic and hadronic decays of $Z$-boson whose branching
ratios are ${\cal B}[Z\to\mu^+\mu^-]\simeq 3.4\%$ and ${\cal B}[Z\to
jj]~70\%$, respectively~\cite{Agashe:2014kda}.
The dimuon channel has limited number of events, but is promised to be
observed because of the clear signal and fine momentum-resolution.
The signal in dijet channel has large number of events because of the 
large branching ratio, but may be affected by large reducible background
events and less momentum resolution for jet measurements.

In the fermion and vector DM models, another Higgs boson $H_2$ has
been introduced.
If $H_2$ can be produced on-shell, and its decay branching ratio into
a DM pair is sizable, we expect the invisible decay of $H_2$ as
a signal of the DM production.
This can be investigated by searching for another peak in the recoil
mass distribution in events with $Z$-boson plus missing energy. 
On the other hand, even if $H_2$ cannot be produced on-shell or $H_2$
cannot decay into a DM pair by kinematical reasons, a DM pair can be
produced through the off-shell propagation of $H_1$ and $H_2$. 
The collider signal for this case would be an excess in a relatively
wide region in the recoil-mass distribution for the events with
$Z$-boson plus missing energy.
As we discussed, the propagators of $H_1$ and $H_2$ are
$\propto|(t-m_{H_1}^2+im_{H_1}\Gamma_{H_1})^{-1} -
(t-m_{H_2}^2+im_{H_2}\Gamma_{H_2})^{-1}|^2$. 
Thus, for $m_{H_1}^2<t<m_{H_2}^2$, this gives a constructive
interference.
We emphasize here again that the mass, spin of DM as well as the mass of
another Higgs boson can be explored by studying the shape of the
recoil-mass distribution.

\subsubsection{Scalar DM, $m_D\ge m_{H_2}/2$ or
   $m_{H_2}\ge\sqrt{s}-m_Z$ cases}

In the cases of the scalar DM model, and the fermion and vector DM
models with $m_D\ge m_{H_2}/2$ or $m_{H_2}\ge\sqrt{s}-m_Z$, a DM pair is
produced via $e^+e^-\to ZH_1^*(/H_2^*)\to ZDD$.
The amplitude is proportional to $\lambda_{HS}$, $\lambda_\chi=y_\chi
s_\alpha c_\alpha$ and $\lambda_V=g_V s_\alpha c_\alpha$ in the scalar,
fermion and vector DM models, respectively.
Therefore by observing the $Z$-boson plus missing energy events at large
$M_{\rm rec}(\ge 2m_D)$, these parameters can be determined or
constrained.
We study the feasibility of detecting this process at the ILC by a
simple MC simulation using MadGraph version 5~{\tt
MadGraph5}~\cite{Alwall:2014hca}. 
The signal of the process can be a reconstructed $Z$-boson plus missing
energy;
\begin{align}
 e^+ e^- \to ZH_1^*(/H_2^*)\to ZDD\to (jj~{\rm or}~\mu^+\mu^-) +
 E,\hspace{-10pt}/ 
\end{align}
where we consider the hadronic and muonic decays of the $Z$-boson.
Major SM background events in the hadronic channel come from 
$e^+e^-\to Z\nu_\ell\bar{\nu}_\ell\to jj\nu_\ell\bar{\nu}_\ell$ with
$\ell=e,\mu,\tau$.
On the other hand, those in the muonic channel come from
$e^+e^-\to \mu^+\mu^-\nu_\ell\bar{\nu}_\ell$ with $\ell=e,\mu,\tau$. 
The total cross sections for these background processes in the dimuon
and dijet channels are calculated to be 107~fb and 345~fb,
respectively, at $\sqrt{s}=500$~GeV.

To reduce the SM background, we consider following kinematical cuts;
\begin{subequations}
 \begin{align}
  &p_T^Z\ge 100~{\rm GeV},\\
  &\left|\eta^Z\right|\le 1,\\ 
  &2m_{D}\le M_{\rm rec}\le 2m_{D}+50~{\rm [GeV]}.
 \end{align}\label{eq:cuts}
\end{subequations}
We simulate the signal and background events by using {\tt MadGraph5}
at the parton level, and estimate the efficiencies by these cuts. 
In Table~\ref{tab:SDM}, we summarize the total cross-section of the
signal process divided by $\lambda_{HS}^2$, the efficiencies for the
signal and background events by kinematical cuts as a function of the DM
mass $m_S$ for $\sqrt{s}=500$~GeV. 
The DM mass is examined from 80~GeV to 160~GeV.
By these kinematical cuts, about 60\% to 20\% of signal events survive
depending on the mass of DM, while background events are suppressed to
${\cal O}(0.1\%)$ level in the dimuon channel and ${\cal O}(1\%)$
level in the dijet channel. 

\begin{figure}[ht]
 \begin{center}
  \includegraphics[width=0.6\textwidth]{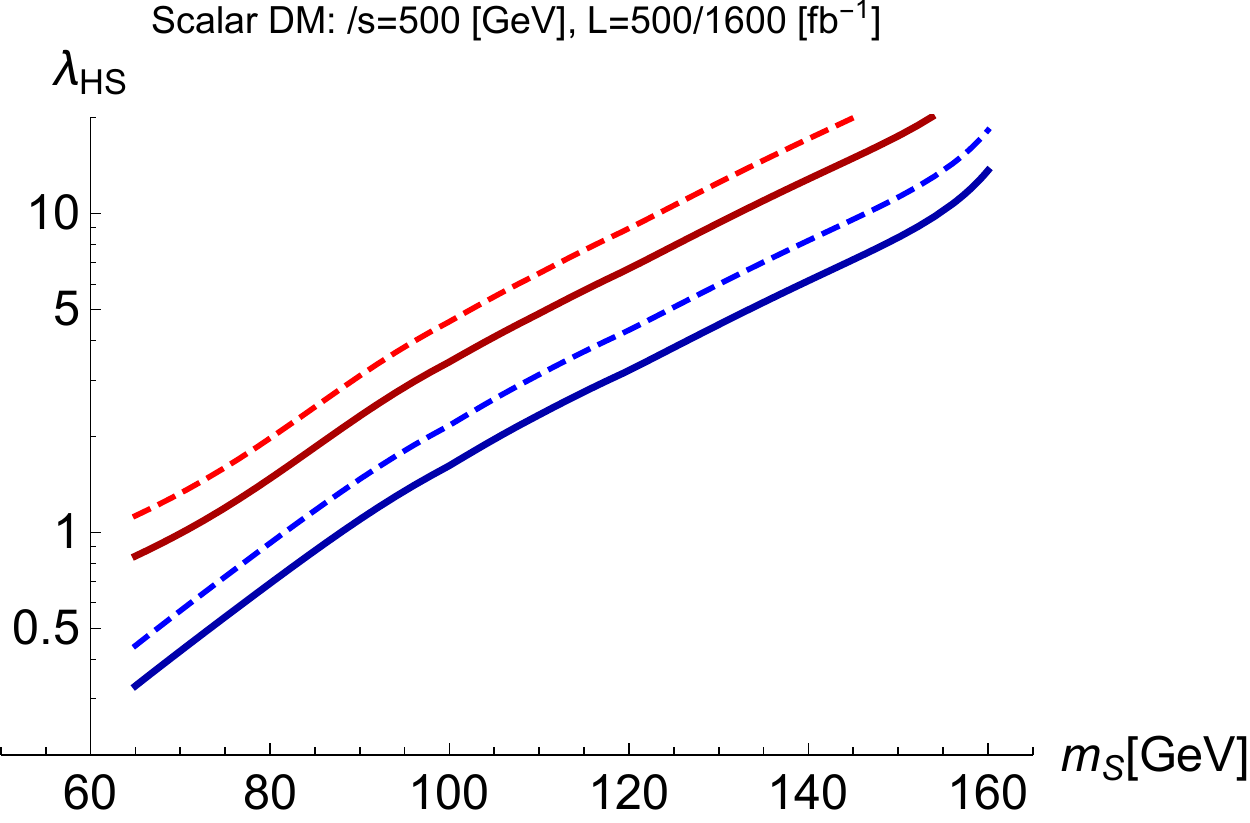}
  \caption{
  Contour plots for the discovery potential at 95\% C.L.\ in $e^+e^-\to
  ZDD$ searches at the ILC with $\sqrt{s}=500$~GeV and ${\mathcal
  L}=500$~(dashed), 1600~(solid)~[fb$^{-1}$].
  Red contours are the limits by using the dimuon channel of $Z$-boson
  decay, and blue contours are the limits by using the dijet channel.
}\label{fig:SDM}
 \end{center}
\end{figure}

We estimate the significance of detecting the excess in events with
$Z$-boson plus missing energy by 
\begin{align}
 S=\frac{\sigma_{ZDD}{\cal B}(Z\to\mu^+\mu^-/jj)\epsilon_S{\cal
 L}}{\sqrt{\sigma_{\rm BG}\epsilon_B{\cal L}}}. 
\end{align}
We say $S\ge 5$ is required to discover signal events.
Because the cross section scales with $\lambda_{HS}^2$, we can evaluate
the lower limit of $\lambda_{HS}$ ($\lambda_{HS}^{\rm min.}$) to be
detected by a certain accumulated luminosity for each $m_S$.
In Table~\ref{tab:SDM}, our estimations for $\lambda_{HS}^{\rm min.}$ are
also listed assuming ${\cal L}=500$~fb$^{-1}$ and 1600~fb$^{-1}$. 
In Fig.~\ref{fig:SDM}, we plot $\lambda^{\rm min.}_{HS}$ in the dimuon
channel (red lines) and the dijet channel (blue lines) for ${\cal
L}=500$~fb$^{-1}$ (dashed lines) and 1600~fb$^{-1}$ (solid lines). 
We find that in the future ILC experiment with $\sqrt{s}=500$~GeV,
$\lambda_{HS}\le 1$ can be surveyed only for light DM
cases~($m_S\lesssim80$~GeV), and only $\lambda_{HS}\ge10$ can be
surveyed for heavier DM cases like $m_S\ge150$~GeV.

\begin{table}[t]
 \begin{center}
  \begin{tabular}{l|ccccccc}
   \multicolumn{8}{c}{Scalar DM Model, $\sqrt{s}=500$~GeV} \\
   $m_S$~[GeV] & 80 & 90 & 100 & 110 & 120 & 140 & 160 \\ \hline \hline
   $\sigma_{ZSS}/\lambda_{HS}^2$~[fb] & 1.37 & 0.69 & 0.38 & 0.22 & 0.13
		       & 0.046 & 0.014 \\
   $\epsilon_S$~[\%] & 56 & 51 & 48 & 47 & 48 & 45 & 23 \\
   \hline \hline
   \multicolumn{1}{c|}{$Z\to\mu^+\mu^-$} &&&&&& \\
   $\epsilon_B$~[\%] & 0.19 & 0.24 & 0.32 & 0.41 &
		       0.55 & 0.80 & 0.44 \\
   $\lambda_{HS}^{\rm min.}$ (${\cal L}=500$~fb$^{-1}$) & 1.97 & 3.09 &
	       4.56 & 6.47 & 8.93 & (17.1)  & (37.5) \\
   $\lambda_{HS}^{\rm min.}$ (${\cal L}=1600$~fb$^{-1}$) & 1.47 & 2.31 &
	       3.41 & 4.84 & 6.68 & (12.8) & (28.1) \\
   \hline \hline
   \multicolumn{1}{c|}{$Z\to jj$} &&&&&& \\
   $\epsilon_B$~[\%] & 1.2 & 1.6 & 2.1 & 2.9 & 3.8 & 5.6 & 3.2 \\
   $\lambda_{HS}^{\rm min.}$ (${\cal L}=500$~fb$^{-1}$) & 0.922 & 1.46 &
	       2.16 & 3.11 & 4.29 & 8.20 & (18.2) \\
   $\lambda_{HS}^{\rm min.}$ (${\cal L}=1600$~fb$^{-1}$) & 0.689 & 1.09
	   & 1.61 & 2.33 & 3.21 & 6.13 & (13.6) \\
  \end{tabular}
  \caption{Signal and background efficiencies for detecting $e^+e^-\to
  ZH_1\to ZDD$ process at the ILC.
  Prospects for parameter constraints are also shown assuming ${\cal
  L}=500$~fb$^{-1}$ and 1600~fb$^{-1}$.
  Limits on the couplings larger than $4\pi$ are presented within parentheses. 
  }\label{tab:SDM}
 \end{center}
\end{table}

\begin{figure}[ht]
 \begin{center}
  \includegraphics[width=0.48\textwidth]{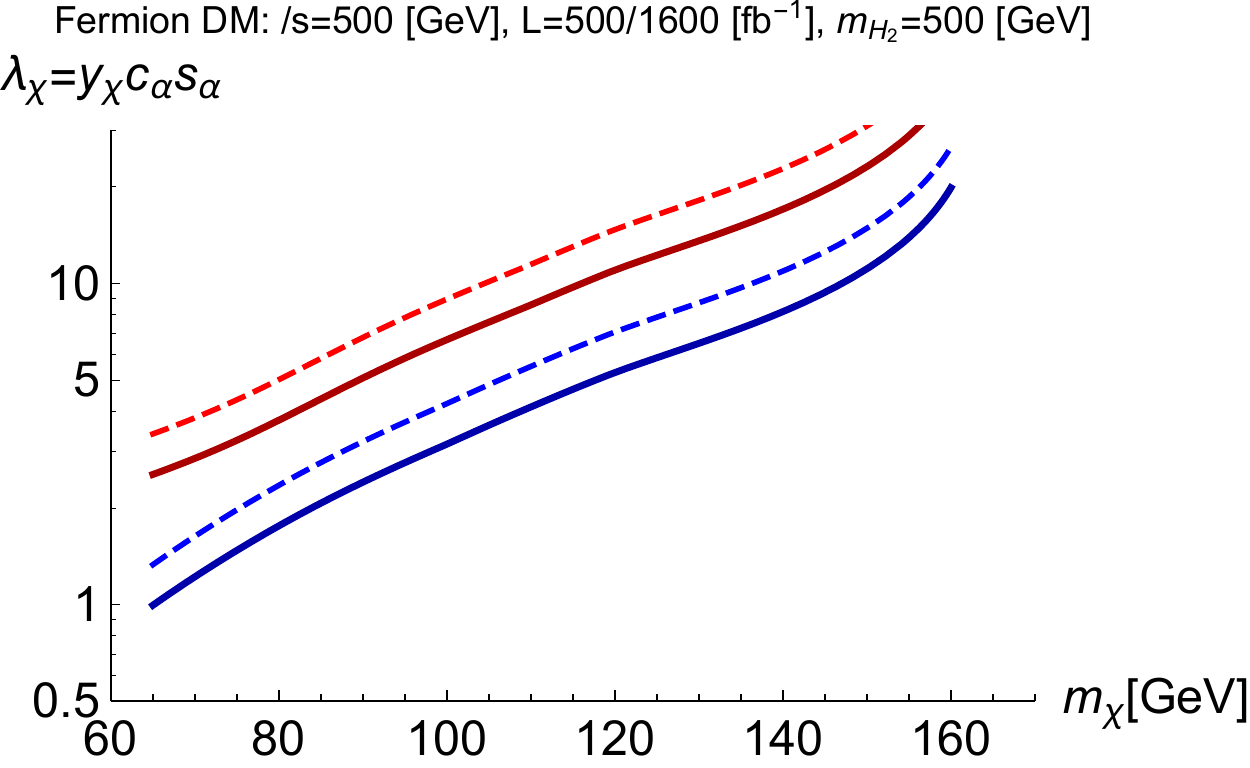}
  \includegraphics[width=0.48\textwidth]{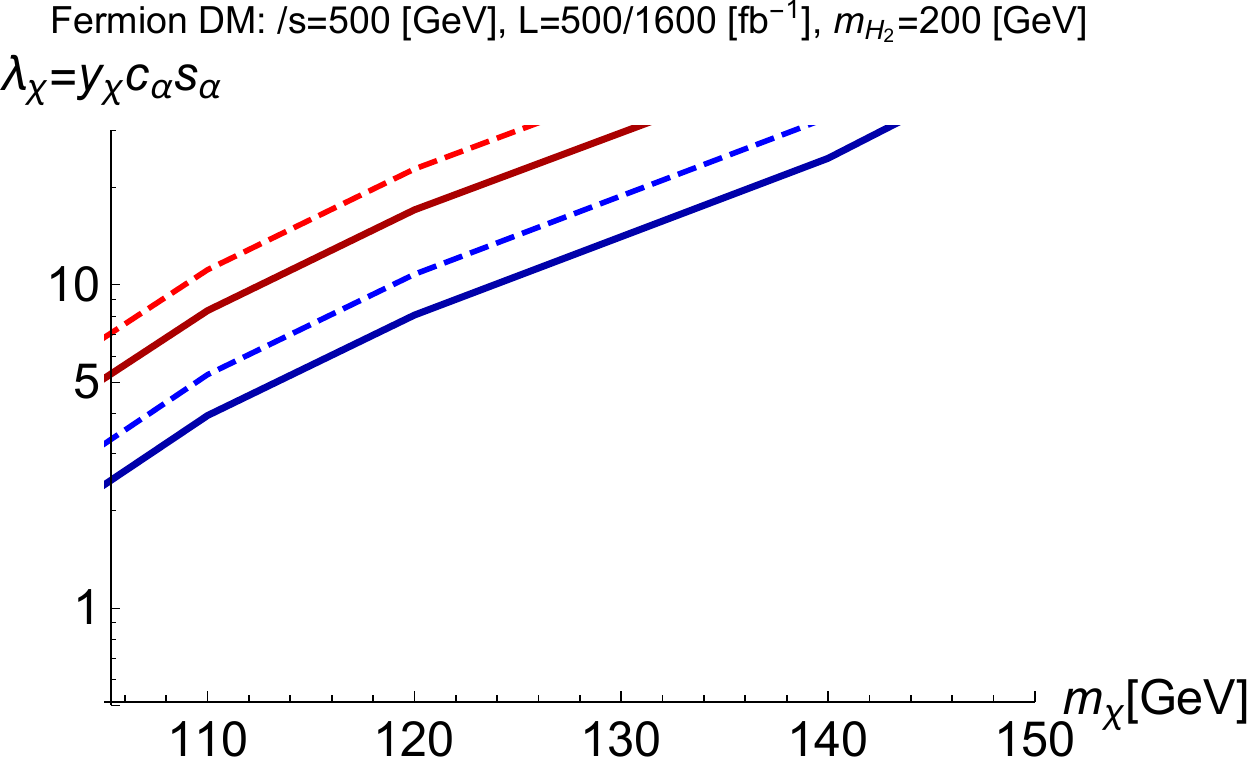}
  \caption{The same figure as Fig.~\ref{fig:SDM}, but for the fermion DM
  model with $m_{H_2}=500$~GeV~[left] and 200~GeV.}\label{fig:FDM} 
 \end{center}
\end{figure}

In the fermion and vector DM models, the production cross section
depends on $m_{H_2}$ as well.
We consider $m_{H_2}=500$~GeV and $m_{H_2}=200$~GeV for example.
In the large $m_{H_2}$ limit, the diagram with $H_2$ propagator
decouples, and the collider phenomenology becomes the same as that for
the simple extension of the SM by adding only fermion or vector
DM.\footnote{%
Modifying the SM Higgs boson couplings by $\kappa_V=\kappa_F=\cos\alpha$
is an important consequence of making the Higgs portal DM models SM
gauge invariant and unitary~\cite{Chpoi:2013wga,Baek:2015lna}.}
In Table~\ref{tab:FDM}, we summarize the analysis for the fermion DM
model with $m_{H_2}=500$~GeV. 
The signal cross section, efficiencies, and the lower limit of
$\lambda_\chi=y_\chi c_\alpha s_\alpha$ to be detected in dimuon and dijet
channels by $5\sigma$ C.L.\ assuming ${\cal L}=500$ and 1600~fb$^{-1}$
are summarized.
In the left panel of Figure~\ref{fig:FDM}, we plot $\lambda_\chi^{\rm
min.}$ in the dimuon channel (red lines) and the dijet channel (blue
lines) for ${\cal L}=500$~fb$^{-1}$ (dashed lines) and 1600~fb$^{-1}$
(solid lines).
We perform the same analysis for $m_{H_2}=200$~GeV, and the results are
shown in the right panel of Figure~\ref{fig:FDM}.

\begin{table}[t]
 \begin{center}
  \begin{tabular}{l|ccccccc}
   \multicolumn{8}{c}{Fermion DM Model, $\sqrt{s}=500$~GeV,
   $m_{H_2}=500$~GeV} \\ 
   $m_\chi$~[GeV] & 80 & 90 & 100 & 110 & 120 & 140 & 160 \\ \hline \hline
   $\sigma_{ZFF}/\lambda_\chi^2$~[fb] & 0.76 & 0.53 & 0.37 & 0.26 & 0.18 &
			   0.077 & 0.025 \\
   $\epsilon_S$~[\%] & 15 & 13 & 13 & 13 & 13 & 15 & 6 \\
   \hline \hline
   \multicolumn{1}{c|}{$Z\to\mu^+\mu^-$} &&&&&&& \\
  $\epsilon_B$~[\%] & 0.19 & 0.24 & 0.32 & 0.41 &
		       0.55 & 0.80 & 0.44 \\
   $\lambda_\chi^{\rm min.}$ (${\cal L}=500$~fb$^{-1}$) & 5.01 & 6.78 &
	       8.92 & 11.4 & (14.7) & (22.7) & (54.8) \\
   $\lambda_\chi^{\rm min.}$ (${\cal L}=1600$~fb$^{-1}$) & 3.75 & 5.07 &
	       6.67 & 8.56 & 11.0 & (17.0) & (41.0) \\
   \hline \hline
   \multicolumn{1}{c|}{$Z\to jj$} &&&&&&& \\
   $\epsilon_B$~[\%] & 1.2 & 1.6 & 2.1 & 2.9 & 3.8 & 5.6 & 3.2 \\
   $\lambda_\chi^{\rm min.}$ (${\cal L}=500$~fb$^{-1}$) & 2.35 & 3.21 &
	       4.22 & 5.51 & 7.06 & 10.9 & (26.5) \\
   $\lambda_\chi^{\rm min.}$ (${\cal L}=1600$~fb$^{-1}$) & 1.76 & 2.40 &
	       3.16 & 4.12 & 5.28 & 8.16 & (19.8) \\
  \end{tabular}
  \caption{The same as Table~\ref{tab:SDM}, but for the fermion DM
  case.}\label{tab:FDM} 
 \end{center}
\end{table}

\begin{figure}[ht]
 \begin{center}
  \includegraphics[width=0.48\textwidth]{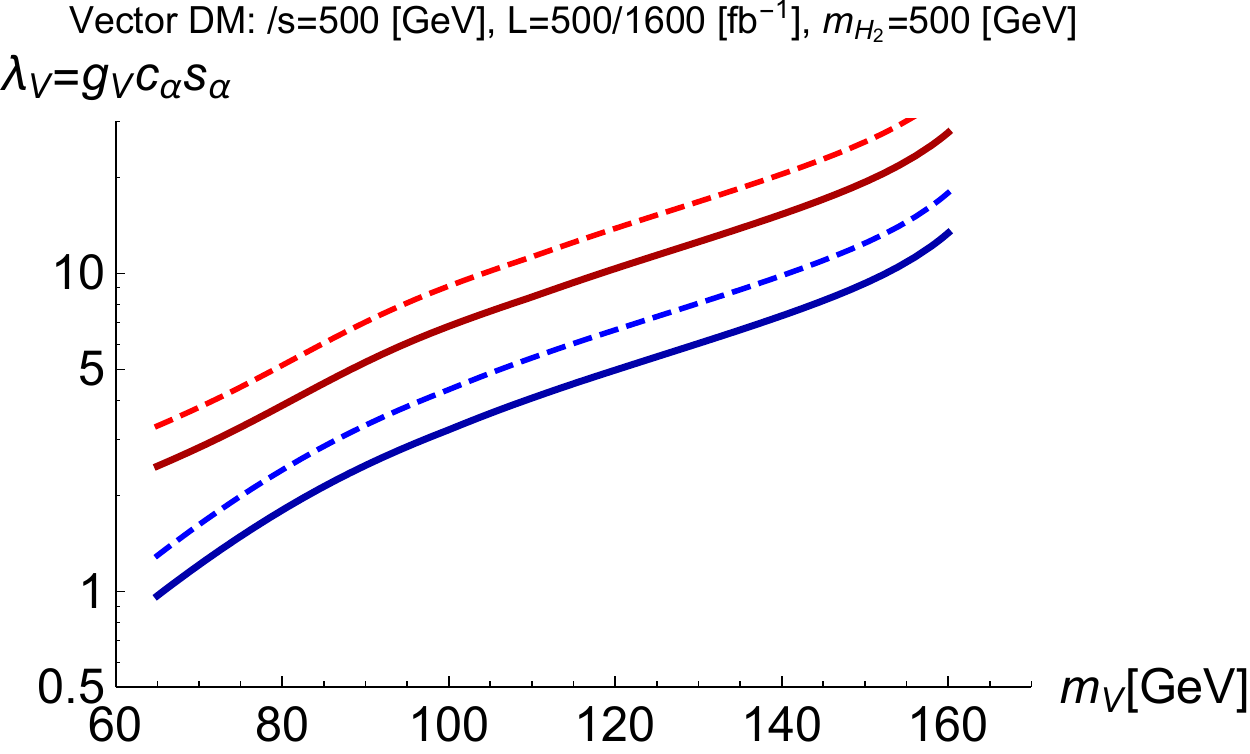}
  \includegraphics[width=0.48\textwidth]{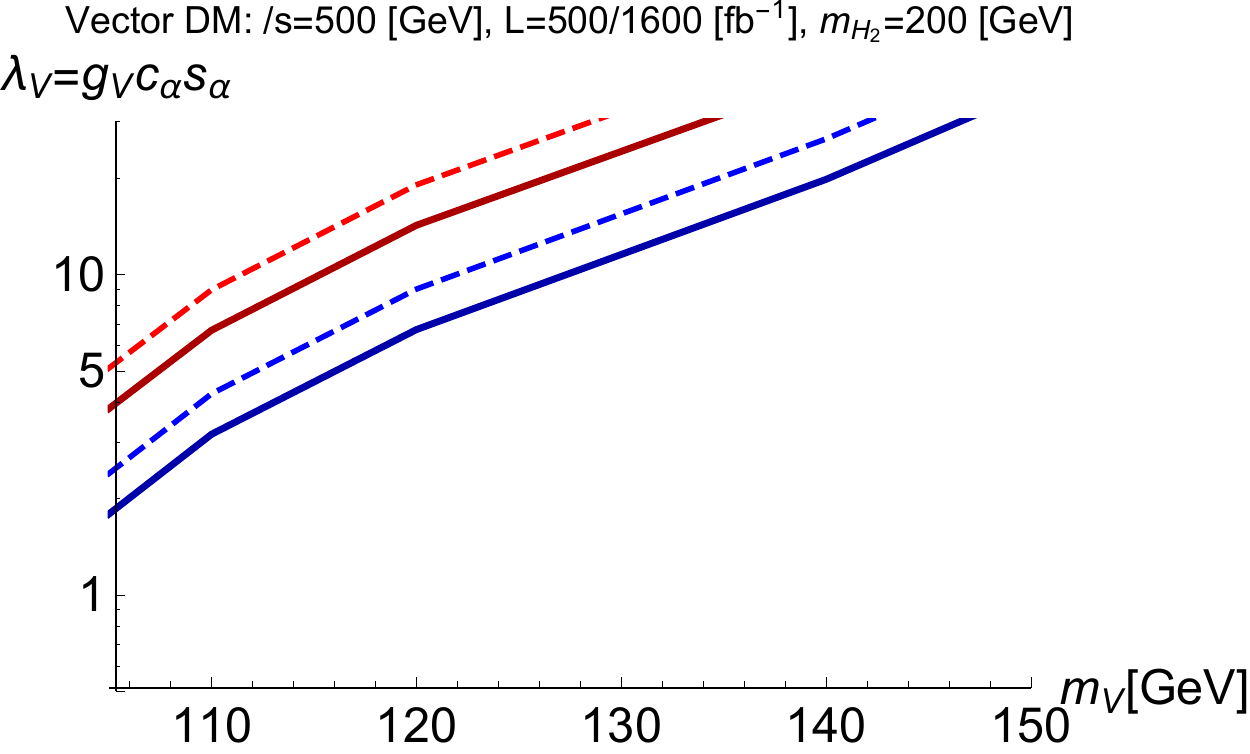}
  \caption{The same figure as Fig.~\ref{fig:SDM}, but for the vector DM
  model with $m_{H_2}=500$~GeV~[left] and 200~GeV.}\label{fig:VDM}
 \end{center}
\end{figure}

In Table~\ref{tab:VDM} and Figure~\ref{fig:VDM}, the same analyses for
the vector DM model are summarized, where $\lambda_V=g_Vc_\alpha
s_\alpha$ is constrained by the measurement.

\begin{table}[t]
 \begin{center}
  \begin{tabular}{l|ccccccc}
   \multicolumn{8}{c}{Vector DM Model, $\sqrt{s}=500$~GeV,
   $m_{H_2}=500$~GeV} \\
   $m_V$~[GeV] & 80 & 90 & 100 & 110 & 120 & 140 & 160 \\ \hline \hline
   $\sigma_{ZVV}/\lambda_{V}^2$~[fb] & 0.74 & 0.47 & 0.31 & 0.21 & 0.14
		       & 0.064 & 0.026 \\
   $\epsilon_S$~[\%] & 15 & 14 & 15 & 16 & 19 & 23 & 13 \\ \hline \hline
   \multicolumn{1}{c|}{$Z\to\mu^+\mu^-$} &&&&&&& \\
   $\epsilon_B$~[\%] & 0.19 & 0.24 & 0.32 & 0.41 & 0.55 & 0.80 & 0.44 \\
   $\lambda_{V}^{\rm min.}$ (${\cal L}=500$~fb$^{-1}$) & 5.13 & 7.02 &
	       9.10 & 11.2 & (13.8) & (20.4) & (37.0) \\
   $\lambda_{V}^{\rm min.}$ (${\cal L}=1600$~fb$^{-1}$) & 3.84 & 5.25 &
	       6.80 & 8.41 & 10.3 & (15.3) & (27.6) \\ \hline \hline
   \multicolumn{1}{c|}{$Z\to jj$} &&&&&&& \\
   $\epsilon_B$~[\%] & 1.2 & 1.6 & 2.1 & 2.9 & 3.8 & 5.6 & 3.2 \\
   $\lambda_{V}^{\rm min.}$ (${\cal L}=500$~fb$^{-1}$) & 2.40 & 3.33 &
	       4.30 & 5.41 & 6.63 & 9.82 & (17.9) \\
   $\lambda_{V}^{\rm min.}$ (${\cal L}=1600$~fb$^{-1}$) & 1.80 & 2.49 &
	       3.22 & 4.05 & 4.96 & 7.34 & (13.4) \\
  \end{tabular}
  \caption{The same as Table~\ref{tab:SDM}, but for the vector DM
  case.}\label{tab:VDM}
 \end{center}
\end{table}

\subsubsection{Fermion and Vector DM models with $m_D\le m_{H_2}/2$}

In the case with $m_D\le m_{H_2}/2$ in the fermion and vector DM models,
if $H_2$ is light enough to be produced, DM can be searched
for as an invisible decay of $H_2$, since the coupling of the dark
matter to the another Higgs boson is expected to be sizable in the Higgs
portal scenario.
Here, we study the production of $H_2$ in $e^+e^-\to ZH_2$ where $Z$
decays into $\mu^+\mu^-$ or $jj$, and $H_2$ decays into $DD$.
This signal can be a part of the inclusive $H_2$ search in $e^+e^-\to
ZX$ process where $H_2$ can be detected by a new peak in the recoil
mass distribution at $M_{\rm rec.}\simeq m_{H_2}$.

\begin{figure}[ht]
 \begin{center}
  \includegraphics[width=0.6\textwidth]{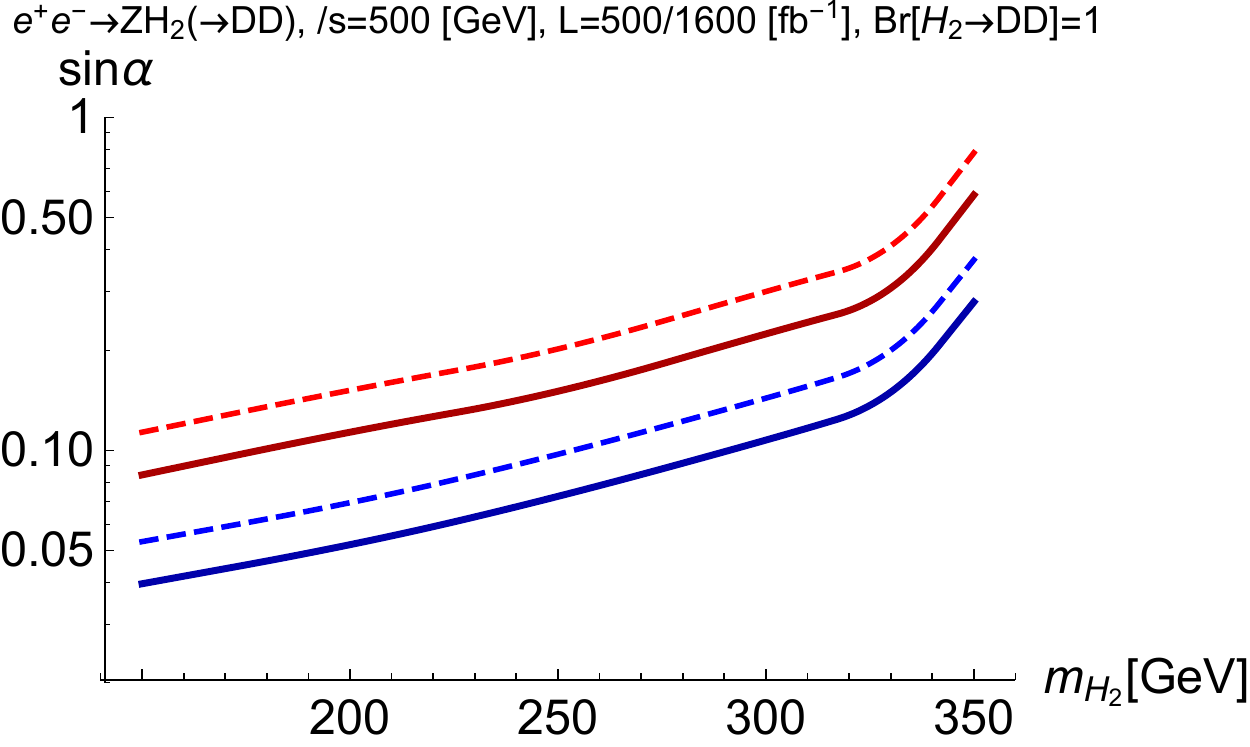}
  \caption{
  Contour plots for the discovery potential at 95\% C.L.\ in $e^+e^-\to
  ZH_2$ searches using the invisible decay mode of $H_2\to DD$ at the
  ILC with $\sqrt{s}=500$~GeV and ${\mathcal L}=500$~(dashed),
  1600~(solid)~[fb$^{-1}$].
  Red contours are the limits by using the dimuon channel of $Z$-boson
  decay, and blue contours are the limits by using the dijet channel.
  ${\rm Br}(H_2\to DD)=1$ is assumed.
  }\label{fig:on}
 \end{center}
\end{figure}

We also perform a simulation analysis to study to what extent the signal
can be detected at future lepton colliders with $\sqrt{s}=500$~GeV.
The total event rate is estimated by $\sigma(ZH_2){\cal B}(H_2\to
DD)$ where $\sigma(ZH_2)$ is proportional to $s_\alpha^2$.
We consider a scenario where ${\cal B}(H_2\to DD)$ is large.
For simplicity, we take ${\cal B}(H_2\to DD)=1$.
Then, $s_\alpha$ is determined or constrained by the experimental
measurement.
To enhance the signal significance in the presence of background events,
we apply the same kinematical cuts in Eqs.~(\ref{eq:cuts}) but the cut
on $M_{\rm rec.}$ is replaced by 
\begin{align}
 \left|M_{\rm rec}-m_{H_2}\right|\le10~[{\rm GeV}], 
\end{align}
because of the sharp peak in the signal events.
In Table~\ref{tab:on}, we summarize the signal cross-section, signal and
background efficiencies by cuts, and the lower limits of $s_\alpha$ to be
observed at $5\sigma$ C.L.\ by using dimuon or dijet decays of $Z$
boson and by assuming ${\cal L}=500$ or 1600~fb$^{-1}$. 
We find that $s_\alpha\simeq 0.1$-0.2 can be investigated for
$m_{H_2}=150$-300~GeV under the assumption of ${\cal B}[H_2\to DD]=1$.
In Fig.~\ref{fig:on}, the lower limits of $s_\alpha$ to be observed are
plotted as a function of $m_{H_2}$.

We make some comments on this analysis.
First, because we have assumed ${\cal B}[H_2\to DD]=1$, the signal
sensitivity may be maximized.
For smaller ${\cal B}[H_2\to DD]$, the number of signal events is
decreased and the sensitivity on $s_\alpha$ would be weakened.
In addition, the analysis does not depend on $m_D$, except demanding
$m_D\le m_{H_2}/2$.
In the actual models, branching ratios of $H_2$ should be predicted and
calculated as a function of $s_\alpha$, $m_{H_2}$, $m_D$ and
$\Gamma(H_2\to H_1H_1)$.
The last quantity can be replaced by model parameters in the Higgs
potential in the model.
More concrete analysis may be required to determine the model parameters
in general situations.
Second, there is no distinction between the fermion DM model and vector
DM model in this measurement, since the recoil mass distribution behaves
just a sharp peak.
Events in the off-peak region may be useful to distinguish the models
based on kinematics, if enough number of events are collected.

\begin{table}[t]
 \begin{center}
  \begin{tabular}{l|ccccc}
   \multicolumn{6}{c}{$e^+e^-\to ZH_2(\to DD)$, $\sqrt{s}=500$~GeV,
   ${\cal B}[H_2\to DD]=1$} \\
   $m_{H_2}$~[GeV] & 150 & 200 & 250 & 300 & 350 \\ \hline \hline
   $\sigma_{ZH_2}/s_\alpha^2$~[fb] & 52.9 & 43.0 & 32.1 & 21.4 & 11.9 \\
   \hline
   $\epsilon_S$~[\%] & 87 & 86 & 85 & 79 & 12 \\
   \hline \hline
   \multicolumn{1}{c|}{$Z\to\mu^+\mu^-$} &&&&&\\
   $\epsilon_B$~[\%] & 0.046 & 0.094 & 0.16 & 0.31 &
		       0.11 \\
   $s_{\alpha}^{\rm min.}$ (${\cal L}=500$~fb$^{-1}$) & 0.11 & 0.15 &
	       0.20 & 0.30 & 0.78 \\
   $s_{\alpha}^{\rm min.}$ (${\cal L}=1600$~fb$^{-1}$) & 0.084 & 0.11 &
	       0.15 & 0.22 & 0.58 \\
   \hline \hline
   \multicolumn{1}{c|}{$Z\to jj$} &&&&&\\
   $\epsilon_B$~[\%] & 0.30 & 0.56 & 1.17 & 2.13 & 0.75 \\
   $s_{\alpha}^{\rm min.}$ (${\cal L}=500$~fb$^{-1}$) & 0.053 & 0.069 &
	       0.097 & 0.14 & 0.37 \\
   $s_{\alpha}^{\rm min.}$ (${\cal L}=1600$~fb$^{-1}$) & 0.040 & 0.052 &
	       0.73 & 0.11 & 0.28 \\
  \end{tabular}
  \caption{Signal and background efficiencies for detecting $e^+e^-\to
  ZH_2$ process at the ILC.
  Prospects for parameter constraints are also shown assuming ${\cal
  L}=500$~fb$^{-1}$ and 1600~fb$^{-1}$.
  }\label{tab:on}
 \end{center}
\end{table}

\subsection{Comparison with Higgs portal DM models within EFT}

As we have seen above, the collider signals of the Higgs portal DM
models, which are UV-completed to preserve the gauge invariance,
renormalizability and unitarity, are more complicated than the simple
EFT-based models.
The presence of the second Higgs boson which is inevitable to make the
models with fermion or vector DM suitable with our requirement gives
characteristic signals for the DM production as well as the new scalar
itself.

For $m_{H_2}\gg \sqrt{s}$, the distribution cannot be distinguished from
that of the EFT calculation. 
Thus, a long tail or a roll in the high-$t$ region does not immediately
imply the unitary violation, but can be regarded as a characteristic
signal of the fermion and vector DM models.
However, the high-energy behavior of the model is completely different
depending on whether the model is renormalizable or not as well as
unitary or not.
Collider phenomenology also depends on the details of the models,
such as presences of new scalars, partners of DM, etc., and also
constraints from the DM relic density, direct detections, etc.
We emphasize that the interplay between these observations has to be
performed in the model with renomalizability and unitarity to combine
the model analyses in different scales.

Before closing this subsection, let us ask when we can ignore the 2nd
scalar propagator in Eq.~(2.29), and use the EFT approach in which
Eq.~(2.30) can be applied. 
Discussion at the ILC is simpler than at the LHC, since the CM energy
$\sqrt{s}$ is fixed at the ILC.
For a fixed $\sqrt{s}$, we can ignore the 2nd scalar propagator in
Eq.~(2.29) if $m_{H_2}^2 \gg \sqrt{s}$. 
Then the effective Higgs portal Lagrangians, Eqs.~(1.2) and (1.3), might
give reasonably good descriptions. 
However it is not true, since the invisible decay width of the 125~GeV
Higgs boson in case of Higgs portal VDM diverges when the VDM mass
approaches zero, which is unphysical~\cite{Baek:2014jga}:\footnote{%
The invisible decay widths in Ref.~\cite{Baek:2014jga} should be
multiplied by $\sin^2 \alpha$ and $\cos^2\alpha$ in order that we get
the physical invisible widths of $H_1$ and $H_2$, respectively.}: 
\begin{equation}
( \Gamma_h^{\rm inv} )_{\rm EFT} 
= \frac{\lambda_{VH}^2}{128 \pi} \frac{v_H^2 m_h^3}{m_V^4} \ 
\left( 1 - \frac{4 m_V^2}{m_h^2} + 12 \frac{m_V^4}{m_h^4} \right) 
\left( 1 - \frac{4 m_V^2}{m_h^2} \right)^{1/2}  
\end{equation}
On the other hand, it is perfectly finite in the full renormalizable and
unitary model, since 
$m_V = g_V v_\varphi /2$~\cite{Baek:2014jga}:
\begin{equation}
\Gamma_h^{\rm inv} = \frac{g_V^2}{32 \pi} \frac{m_h^3}{m_V^2} 
\  \left( 1 - \frac{4 m_V^2}{m_h^2} 
+ 12 \frac{m_V^4}{m_h^4} \right) \left( 1 - \frac{4 m_V^2}{m_h^2} \right)^{1/2} 
 \sin^2 \alpha
\end{equation}
Note that there is more parameter, $\alpha$, in Eq.~(4.8), compared with
Eq.~(4.7) for  
the invisible decay width in the VDM EFT with Higgs portal.
For massive VDM, $v_\varphi \neq 0$ so that Eq.~(4.8) never diverges
when $m_V$ becomes very light in the limit $g_V \rightarrow 0$. 
From the usual EFT view point, Eq.~(1.3) should be good at low energy as
long as $m_{H_2} \gg m_{H_1}$, which however is not the case for the Higgs
invisible decay width.  It is not clear {\it a priori}  when and where
the EFT descriptions would fail in this particular physical quantity.
Based on this example, it would be safer to work in the minimal
renormalizable and unitary models for fermion and vector DM with Higgs
portal.

\section{Conclusion}

In this paper, we have performed the detailed study of Higgs portal
scalar, fermion and vector DM models at the ILC.
We consider the renormalizable, unitary and gauge invariant models, and
compare the results with those obtained within the effective field
theories for the Higgs portal fermion and vector DM models.
For the singlet fermion and vector DM cases, the force mediator involves
two scalar propagators, the SM-like Higgs boson and the dark Higgs boson.
We have shown that their interference generates interesting and
important patterns in the mono-$Z$ plus missing $E_T$ signatures at the
ILC, and the results are completely different from those obtained within
the EFT.
Compared with the EFT, our models have at least two extra parameters,
the singlet-like scalar mass $m_{H_2}$ and the scalar mixing angle
$\alpha$.
These parameters are largely unknown yet, except that
$\sin\alpha\lesssim0.53$ from the current LHC data.
The mixing angle $\alpha$ can be probed at an accuracy of ${\cal
O}(1\%)$ or better by precision measurement of the Higgs boson couplings
at the ILC.
By observing the DM pair production and the effects of the
second Higgs boson to it, model parameters can be further constrained.

In addition, as we have shown in Sec.~3, it would be possible to 
distinguish the spin of DM in the Higgs portal scenarios, if the shape
of the recoil-mass distribution could be  observed.
To achieve this, one needs large number of events and careful treatment
of signal and background analysis, thus it is more difficult than
finding the evidence and measuring the masses. 
However, we emphasize this possibility as an theoretical concept.
Otherwise it would be extremely difficult to distinguish them.
Detailed simulation analysis for the significance of separating
different spin ansatz is beyond the scope of this paper.
An analysis at the LHC in the same philosophy will be addressed in a
separate publication~\cite{lhc}.
\section*{Acknowledgements}
We are grateful to Seungwon Baek, Bhaskar Dutta, Tathagata Ghosh and
Teruki Kamon, Alexander Natale, Myeonghun Park, Wan-Il Park and Chaehyun
Yu for discussions on the topics discussed in this paper.  
This work is supported in part by National Research Foundation of Korea
(NRF) Research Grant NRF-2015R1A2A1A05001869, and by the NRF grant
funded by the Korea government (MSIP) (No.\ 2009-0083526) through Korea
Neutrino Research Center at Seoul National University (P.K.). 


\end{document}